\documentclass[conference]{IEEEtran}
\usepackage{graphicx}
\usepackage{caption}
\usepackage{hyperref}
\usepackage{subfigure}
\usepackage{algorithmic}
\usepackage[ruled]{algorithm2e}
\usepackage{amsmath}
\usepackage{xcolor}
\usepackage{ulem}

\ifCLASSINFOpdf
\else
\fi
\makeatletter
\newcommand{\linebreakand}{%
  \end{@IEEEauthorhalign}
  \hfill\mbox{}\par
  \mbox{}\hfill\begin{@IEEEauthorhalign}
}
\makeatother

\begin{document}
%
\title{LESS: Efficient Log Storage System Based on Learned Model and Minimum Attribute Tree}

\author{\IEEEauthorblockN{Zhiyang Cheng*}
\IEEEauthorblockA{School of Software, Tsinghua University\\
chengzy22@mails.tsinghua.edu.cn}
\and
\IEEEauthorblockN{Zizhen Zhu*}
\IEEEauthorblockA{School of Software, Tsinghua University\\
zzz23@mails.tsinghua.edu.cn}
\and
\IEEEauthorblockN{Haoran Dang*}
\IEEEauthorblockA{School of Software, Tsinghua University\\
danghr23@mails.tsinghua.edu.cn}
\linebreakand
\IEEEauthorblockN{Hai Wan}
\IEEEauthorblockA{School of Software, Tsinghua University\\
wanhai@tsinghua.edu.cn}
\and
\IEEEauthorblockN{Xibin Zhao}
\IEEEauthorblockA{School of Software, Tsinghua University\\
zxb@tsinghua.edu.cn}}

\IEEEoverridecommandlockouts
\makeatletter\def\@IEEEpubidpullup{3.5\baselineskip}\makeatother
\IEEEpubid{\parbox{\columnwidth}{
    \noindent\rule{\columnwidth}{0.4pt}\\ 
    *Equal Contribution.
}
\hspace{\columnsep}\makebox[\columnwidth]{}}

\maketitle


\begin{abstract}
In recent years, cyber attacks have become increasingly sophisticated and persistent. Detection and investigation based on the provenance graph can effectively mitigate cyber intrusion. However, in the long time span of defenses, the sheer size of the provenance graph will pose significant challenges to the storage systems. Faced with long-term storage tasks, existing methods are unable to simultaneously achieve lossless information, efficient compression, and fast query support. In this paper, we propose a novel provenance graph storage system, LESS, which consumes smaller storage space and supports faster storage and queries compared to current approaches. We innovatively partition the provenance graph into two distinct components, the graph structure and attribute, and store them separately. Based on their respective characteristics, we devise two appropriate storage schemes: the provenance graph structure storage method based on machine learning and the use of the minimal spanning tree to store the graph attributes. Compared with the state-of-the-art approach, LEONARD, LESS reduces 6.29$\times$ in storage time, while also achieving a 5.24$\times$ reduction in disk usage and an 18.3$\times$ faster query speed while using only 11.5\% of the memory on DARPA TC dataset.
\end{abstract}


%

\section{Introduction}

 



Recently, advanced persistent threats (APT) have become increasingly frequent and intricate. Unlike traditional cyber attacks, APTs can remain undetected with high levels of concealment and lurk for months or even years, posing significant and devastating challenges to cyber security. These carefully planned attacks are executed by professional adversaries using advanced tactics to exploit vulnerabilities in systems and gain access to targeted networks. The APTs can persist for an extended time without triggering alerts until they reach the sensitive information and conceal harmful effects \cite{APTs}. To prevent and mitigate APT attacks, security entities should leverage comprehensive system monitoring and record system execution and events in log files. Auditing logs, known as post-mortem forensic methods for event analysis and investigation, provide primitive evidence of activities in operating systems and the consequently triggered events.

\subsection{Background}
Security entities utilize popular auditing frameworks, such as Linux Audit framework\cite{linuxaudit}, Windows ETW (Event Tracing for Windows)\cite{etw} and FreeBSD's DTrace\cite{dtrace}, to faithfully capture all the system calls of system objects, for example, a process of reading and writing to files. System defenders can perform diagnostic analysis on logs based on the defense systems, such as intrusion detection systems (IDS) and security information and event management (SIEM). However, cyber attacks dwell for a very long time, and it also takes a considerable amount of time to complete the investigation of the attack since getting a POI (Point-Of-Interest) event. For instance, malware can lurk on a target system for more than 365 days\cite{holmes}.
As there is an increasing delay between the initial undetected intrusion and the subsequent triggered events, a growing volume of logs needs to be persistently recorded. Furthermore, in enterprises, the number of machines and devices is increasing every year, and so is the number of devices that need to be logged, putting log data storage a huge financial burden on the organizations. A few years ago, a commercial bank with 200,000 hosts could generate 70PB of data each year\cite{cprpcar}. Therefore, the growing amount of recorded log information and the compelling commercial log storage requirement suggest that we need an efficient solution to support massive log storage and data processing capability.

\begin{figure}[htbp]
  \centering
  \includegraphics[width=\linewidth]{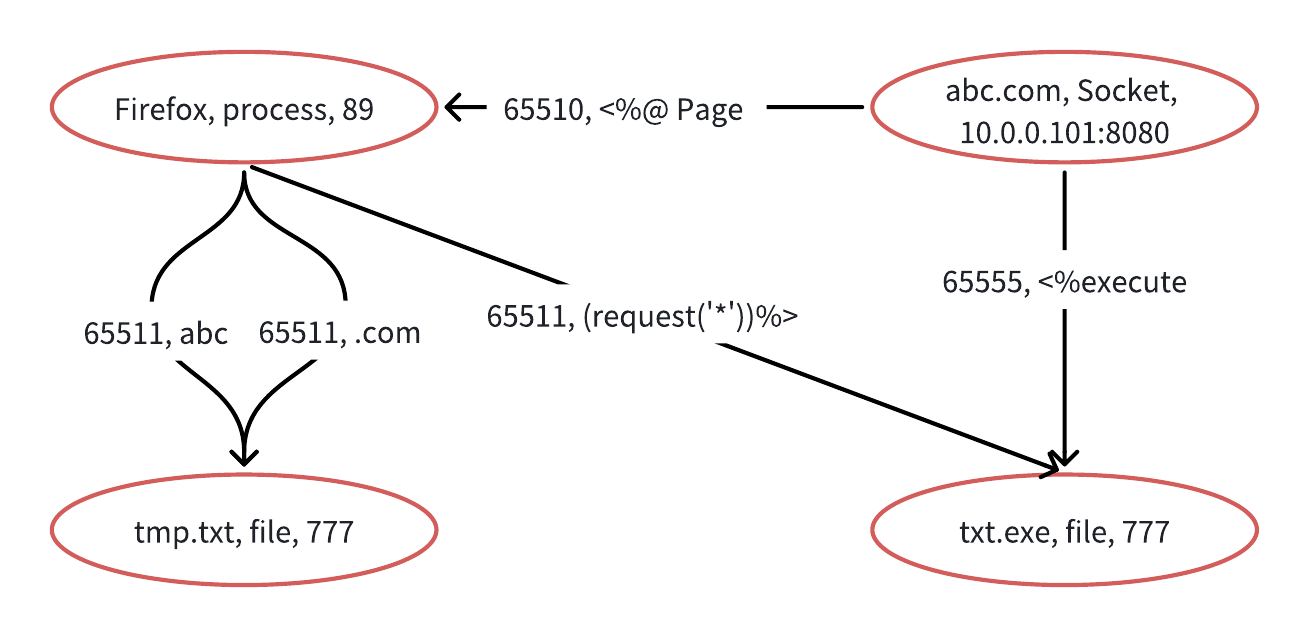}
  \caption{A Provenance Graph Example.}
  \label{provenancegraph}
\end{figure}

Logs are usually semi-structural and non-structural text files. Due to these characteristics, analyzing such logs to detect anomalies in a short time is difficult and error-prone. To address this challenge, provenance graphs are constructed to explicitly describe the interactions between system objects based on log information\cite{backtracking}. The provenance graph directly represents the dependencies and data flow of system objects, also known as causality graph\cite{seal} or dependency graph\cite{nodemerge}. It uses system subjects (e.g., process and threads) and objects (e.g., files, sockets, etc.) as nodes, accesses from system subjects to system objects as edges (e.g., processes writing to files, etc.), and attributes of system objects or operations as properties of the nodes or edges (e.g., process IDs, timestamps, etc.). An example of a provenance graph is illustrated in Figure \ref{provenancegraph}. Using the provenance graph can simplify the intrusion detection and attack investigation\cite{threatdetectionsurvey}. Intrusion detection searches for edges or nodes related to attacks on the provenance graph, creating an entry for attack investigation. Attack investigation, also known as attack provenance, is based on anomalous nodes or edges to determine the initial incursion of an attack (backtrace queries) and identify the consequences of an attack (forward-trace queries) to explain the history of an attack. After processing logs into provenance graphs, the long-term storage of voluminous provenance graphs needs a compelling solution.

In order to achieve long-term storage of provenance graphs, the storage method should support the following features:

\begin{itemize}

\item {{\bfseries Lossless Compression. }The provenance graph storage algorithm needs to reduce physical disk usage while ensuring no loss of information. Maintaining a comprehensive log of all information plays a crucial part in providing complete evidence for subsequent detection and investigation. Any loss of fact records during an attack can impede the analysis from reconstructing the entire attack story,  thereby compromising security defenses.}


\item {{\bfseries Query Support. }The provenance graph storage scheme should facilitate fast reverse operations, i.e., support fast queries. 
It is essential to strike a balance between space and time overheads. While compression tools like Gzip can significantly reduce file size, they come with high computational costs and time overheads during compression and decompression.}
\end{itemize}

While fulfilling the above requirements, the provenance storage method should also exhibit high efficiency in terms of fast storage completion, high compression ratio, and rapid query speed. Efficient processing of the provenance graph into a final file allows storage systems to handle parallel scenarios effectively. Minimizing the size of stored files helps conserve valuable storage space consumption. Fast query speed ensures low latency in detection and investigation processes, facilitating efficient analysis. Therefore, compression methods need to be designed to support simultaneous efficient storage, lossless compression, and fast querying.

\subsection{Existing Work}
Existing work can be divided into three types: storage of provenance graph based on database or compression tool, lossy reduction (discard part of graph information) and lossless compression (retain all graph information).

{\bfseries Database \& Zip. }Prior work uses databases to store the provenance graph. The provenance graph can be processed into a file format, such as a JSON format, and then stored by conventional databases, such as relational databases\cite{mysql}, graph databases\cite{neo4j}, and key-value databases\cite{redis}. The deficiencies of these database systems require additional processing before storage or querying. One of the obstacles to provenance graph storage is its large volume size of data. However, the existing database systems have no targeted processing to reduce massive disk usage, leading to increased system overhead during the graph storage and queries. To address this issue, compression tools, such as Gzip\cite{gzip}, can be used to compress an entire provenance graph into a single file, helping to manage the growing storage requirements to some extent. Gzip is utilized for sequential text compression. However, it exhibits poor query performance in graph structures. Additionally, the expansive decompression overhead significantly impacts the quality of provenance analysis.

{\bfseries Lossy reduction. }Log reduction methods come in handy for performance predicaments when dealing with large amounts of data in provenance graph-based anomaly detection and queries. Log reduction techniques, semantic pruning, information flow preservation, and causal approximation\cite{sok} have been proposed to reduce data in provenance graphs. Semantic pruning removes elements of the provenance graph that do not affect the investigation and detection tasks, e.g., removing nodes and edges representing a process reading and writing to temporary files; causal approximation combines similar or identical substructures of provenance graphs into the same representation; information flow preservation uses simplified structure, e.g., information flow, to replace the original provenance graph or sub-graphs. 
These methods result in information loss, and data invalidation may happen if these methods are utilized to compress a provenance graph before it is stored. These methods are carried out for the purpose of subsequent downstream detection and investigation work but are not designed or optimized for querying. If a provenance graph is processed using the above methods and then stored, the query speed of the compressed graph shows no significant improvement compared to that of the original graph.

{\bfseries Lossless compression. }In addition to lossy reduction, lossless compression methods can support long-term provenance graph storage tasks. Fei et al. develop storage-efficient analysis on enterprise logs (SEAL), which supports query-friendly (QFC)\cite{seal}. For graph structures, SEAL designs merge patterns, where edges and nodes of subgraphs with common information are combined into one new edge and node with new attributes. For graph attributes, Delta coding is utilized to compress long timestamp fields sharing the same prefix, and Golomb coding is applied to express the long integer to a shorter representation. SEAL method mainly considers the merging of structures in the local graph and compressing attributes for fields, which relies on the structure of the original provenance graph and incurs the overhead of decompression in the query. Ma et al.'s LEONARD introduces deep neural networks (DNNs) to store the provenance graph\cite{leonard}. Firstly, the provenance graph is decoupled to convert the graph structure into a sequential data representation, which enables the training of the DNN model, LSTM. As it is difficult to fit a learning model to the training data set with high accuracy, LEONARD designs a calibration table to correct the mispredicted data, specifying the location of misprediction and calibrating the characters. LEONARD utilizes the LSTM to iteratively predict subsequent nodes and edges and correct mispredictions to achieve the query capability. With DNN models and several tables, LEONARD can store the provenance graph with a low disk size usage. However, training a model to learn an entire provenance graph leads to high time costs and low training accuracy. The more mispredictions need to be fixed, the larger the size of the calibration table occupies, affecting the query efficiency.

\subsection{Motivation}
Our design is driven by the following key insights:

First, we have several observations on the limitations of previous graph processing methods, many of which focus solely on either processing the provenance graph structure or compressing attributes to reduce data volume. (1) Previous approaches to process the provenance graph structure typically rely on certain patterns to simplify the graph structure, such as replacing original subgraphs with new nodes or subgraphs\cite{nodemerge, seal, depcomm}. These approaches require that only provenance graphs matching specific patterns can be compressed, limiting the method's applicability to all subgraphs. For example, if the template design is related to the attribute fields of nodes or edges in the provenance graph, it limits the method's applicability to provenance graphs from log sources that do not record these fields. (2) If only attributes are processed, excessive compression of attributes, which constitute the main portion of the provenance graph, can lead to significant decompression overhead during querying\cite{leonard}. Thus, if we handle only the graph structure or graph attributes, we will struggle to achieve both optimal compression rates and fast query speed. Hence, two separate algorithms can be designed to process the provenance graph structure and attributes respectively.

Second, the structure and attributes of provenance graphs exhibit significant differences, necessitating the design of distinct compression algorithms tailored to these two components. The graph structure, which represents the connections between nodes, is typically expressed as adjacency lists or adjacency matrices. Different lists or rows in the adjacency matrix generally lack significant similarity. Using conventional algorithms to extract similarities of the graph structure is difficult. To address this challenge, encoding methods can be employed to represent adjacency lists as shorter vectors. Furthermore, small models can be utilized to store large vectors, similar to using a few prompts to generate a complete text. This requires transforming the graph structure into data suitable for model training, as well as identifying a small model that can achieve high accuracy on the training dataset in a short time. Therefore, we choose the machine learning model XGBoost to store the provenance graph structure(\hyperref[Graph Structure Storage]{\S \uppercase\expandafter{\romannumeral2}-B}).

Graph attributes, which describe information about each node or edge, are typically represented as sets of strings. Attribute entries within the graph often exhibit high similarity because all attribute entries are generated from the same set of rules, and most logs represent repetitive activities of the same task. The key fields in the key-value pairs that express information in different attribute entries are usually similar, and the order of these key-value pairs tends to be consistent within the same file. Different attributes related to the same event often share most of their key-value pairs. For graph attributes, algorithms should be designed to extract the frequently occurring common parts among different attribute elements and store them only once to reduce storage space. Therefore, we need a tree data structure to store graph attributes, where a common parent node can represent the shared information of its child nodes. This way, child nodes only need to store the different parts of the attribute entries, avoiding the storage of redundant information already captured by the parent node. Additionally, the fast search characteristics of the tree structure facilitate rapid attribute queries.

\subsection{Our Approach}
In this paper, we present LESS, an efficient provenance graph storage system. We creatively divide the provenance graph into two parts, graph structure and attribute, and store them separately.  Base on the distinct characteristics of graph structure and attribute,  we design respective storage schemes. We utilize a machine learning model, XGBoost, to store the provenance graph structures and construct an approximate minimum spanning tree to store the attributes of nodes and edges of the provenance graph. Our method uses XGBoost model to learn parts of the graph information and quickly completes the graph structure storage. LESS also leverages the locality of logs(\hyperref[Graph Attribute Storage]{\S \uppercase\expandafter{\romannumeral2}-C}) to compress similar log attributes and achieve a higher compression ratio than existing methods. Additionally, LESS enables faster queries by making predictions only once in queries, which is different from previous iterative queries\cite{leonard}. 

We implement the above algorithms in LESS and evaluate it on 3 datasets. Compared to the state-of-the-art approach, LESS outperforms SOTA by 6.29$\times$ faster in storage time while simultaneously reducing the disk storage by 5.24$\times$ and achieving a 18.3$\times$ faster query speed while using only 11.5\% of the memory on the DARPA TC dataset\cite{darpatc}. 

This paper makes the following contributions:

\begin{itemize}

\item {We propose a novel method for log storage that facilitates efficient query execution. Our scheme achieves lossless compression of the provenance graph, outperforming current research methods in compression ratio without sacrificing any information. Our work is particularly suitable for storing provenance graphs over extended periods of defenses of APT. In addition, we adopt targeted storage designs for different parts of the provenance graphs (graph structure and attribute) to minimize storage time cost and support faster query speed. }

\item {We develop a machine learning-based storage system specifically designed for the provenance graph structure. We convert the provenance graph structure data into vectors for the machine learning model to train and store. Our model is feasible to handle large-scale provenance graph data, representing an improvement over existing storage solutions.}

\item {We study the temporal and spatial similarity possessed by logs, which we define as the locality of logs. By leveraging the locality of logs, we develop a compression method for provenance graph attributes. The algorithm calculates the locality of logs to reduce redundancy across an entire provenance graph, resulting in a higher compression ratio compared to processing local nodes or edges.}

\end{itemize}

\section{METHODOLOGY}

\begin{figure*}[!ht]
  \centering
  \includegraphics[width=\linewidth, trim=12mm 215mm 372mm 7mm, clip]{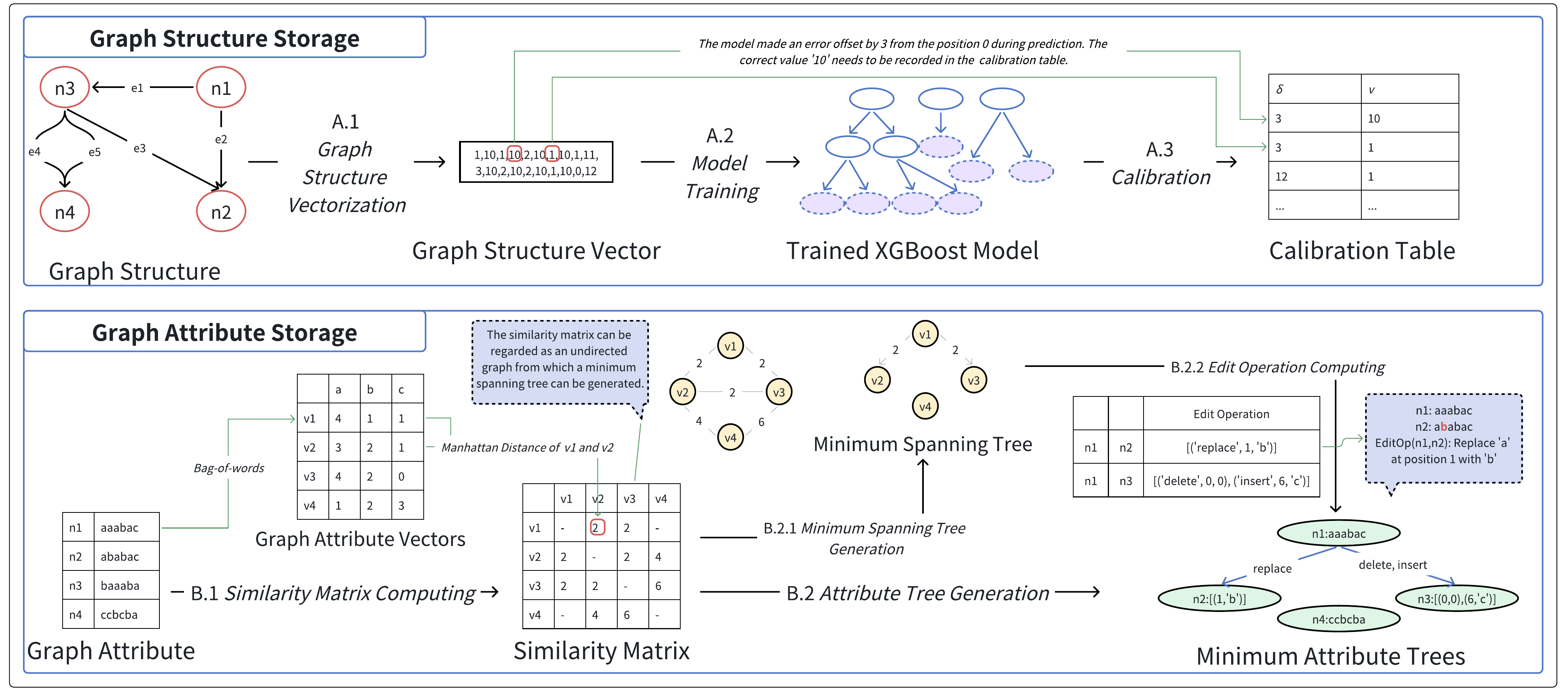}
  \caption{Overview of storage process of LESS: We input the provenance graph in Figure \ref{provenancegraph} to LESS. First, we split the provenance graph into graph structure and graph attribute, and then store them separately. In the structure storage, we process the graph structure into vector (A.1) for training an XGBoost model (A.2). Based on the outputs of XGBoost, a calibration table is generated (A.3). In the attribute storage, we firstly process the node attributes into vectors using the Bag-of-words method, and then calculate the Manhattan distance between two vectors to obtain a similarity matrix with a window size of 3 (B.1). For example, the similarity matrix element 2 represents $d(v1,v2)=\mid4-3\mid+\mid1-2\mid+\mid1-1\mid=2$. The similarity matrix can be viewed as a representation of an undirected graph, where setting the maximum distance to 3 yields the minimum spanning tree (B.2.1), representing the final attribute tree structure. Finally, the corresponding edit operations are generated based on the tree structure to obtain the minimum attribute tree (B.2.2). The trained model, calibration table, and minimum attribute trees are the final outputs. (To simplify the illustration, the node attributes in Figure \ref{provenancegraph} are simplified. Since the storage process for edges is the same as that for nodes, we do not repeat the edge storage process here.)}
  \label{LESSoverview}
\end{figure*}

\subsection{Overview}

LESS takes the provenance graph as input and decouples the graph connection into two parts: graph structures and graph attributes. Separate storage schemes are designed for each part to generate a final file for storage, as illustrated in Figure\ref{LESSoverview}.

The storage of graph structures exclusively processes the structural relations represented by node and edge IDs (\hyperref[Graph Structure Storage]{\S \uppercase\expandafter{\romannumeral2}-B}). Initially, the adjacency list-based graph structure is transformed into a vector record for model training (A.1 Graph Structure Vectorization), wherein we utilize coding techniques to compress data. Subsequently, we employ the XGBoost model (A.2 Model Training) to learn the training dataset consisting of graph structure vectors until the training model predicts the vectors. Finally, based on the outputs of the XGBoost model and the given vectors, a calibration table is constructed (A.3 Calibration) to record and fix the prediction errors.

For the storage of provenance graph attributes, we process the attributes of nodes and edges to construct a minimal attribute tree (\hyperref[Graph Attribute Storage]{\S \uppercase\expandafter{\romannumeral2}-C}). Initially, we compute a similarity matrix (B.1 Similarity Matrix Computing) based on the locality of logs, which represents the similarity between a graph attribute string and its neighboring attribute strings. To expedite the similarity computation, we convert the textual representation of attributes into graph attribute vectors. Computing the Manhattan distance with graph attribute vectors is more efficient than directly calculating the Edit Distance using strings, which also results in higher and more stable accuracy. A similarity matrix can be viewed as a representation of an undirected graph. Based on the attribute similarity matrix, we can compute a minimum spanning tree. By inserting the root node's attributes and the edit operations to all child nodes into the minimum spanning tree, we obtain the minimum attribute tree (B.2 Attribute Tree Generation).

After storing the structure and attributes of the provenance graph, the final artifacts of our storage system, LESS, contain a trained XGBoost model, a calibration table, and minimum attribute trees. Leveraging these files, LESS provides an interface to implement forward-tracing and backward queries, i.e., searching for details of $k$ descendant nodes or ancestors $k$ and connected edges on the original provenance graph (\hyperref[Query]{\S \uppercase\expandafter{\romannumeral2}-D}).


The input of LESS is provenance graphs, and the output is the storable artifacts. When LESS is working with incremental data, the incremental data can be in the form of a provenance graph consisting of streaming log blocks. The incremental provenance graph needs to re-execute the whole storage procedure of LESS, including the structure storage and the attribute storage. LESS requires retraining the model, but the graph structure is usually very small in the total amount of the provenance graph data, so training is very fast and not the main time cost of LESS storage. This is also demonstrated by our experiments (\hyperref[Storage Time Costs]{\S \uppercase\expandafter{\romannumeral3}-B}).

\subsection{Graph Structure Storage}\label{Graph Structure Storage}

{\bfseries Structure Storage Preliminaries.}
We use machine learning models to store graph structures. Previously, there have been efforts to utilize DNNs for storage, achieving promising results\cite{leonard}. DNNs are multi-layered functions which use gradient descent algorithm to optimize the weights of DNNs by training and learn the data distribution. A trained DNN model is able to describe a complex function, which is of poor explainability usually. DNNs can use limited inputs to generate a larger scale of data, such as composing an article with a few prompts. Compared to deep learning models, machine learning models excel in storage tasks, wherein the models output data on the training set without the need to generalize to new instances. (1) Compared to deep learning models, machine learning models are smaller in size and do not have as many weights as the DNN models, which allows for higher compression ratios during storage. (2) Machine learning models converge faster with the same training set size. To achieve the same accuracy on the training set, machine learning requires less data. Therefore, the machine learning model can complete storage tasks in a more time-efficient manner. (3) Machine learning models compute faster, leading to efficient queries for a storage system. We will study the comparison experiments of different models in section \hyperref[differentmodelssec]{\uppercase\expandafter{\romannumeral3}-C-4}. 

We select XGBoost (Extreme Gradient Boosting) as the machine learning model to store the graph structures. The XGBoost model is an optimization of the gradient-boosted decision tree (GBDT) model\cite{xgboost}, which incorporates the boosting enhancement strategy and uses a forward distribution algorithm for greedy learning. XGBoost model has some intuitive advantages in storage over other machine learning models. The XGBoost model size can be conveniently controlled via the hyperparameters max\_depth (the depth of the tree) and n\_estimators (the number of trees). (2) XGBoost can make full use of multi\-core CPU and GPU to accelerate the training and inference process and adapt to the deployment requirements of different hardware devices. (3) Redundant nodes can easily be eliminated in XGBoost after training, further reducing the complexity of the model. The choice of model is not unique. In the LESS framework, it is very convenient to change other models, and do not need to change other module Settings. When deployed in the field, the model can be selected based on the hardware environment and user habits.

During the storage construction, XGBoost learns vectors that represent the graph structure, with each dimension of the vector represented by a number. During querying, the XGBoost model predicts the classification of each number in the original vector. Since it is very difficult for the model to achieve accuracy up to 100\%, the outputs of the model require a calibration table to fix prediction errors. The size of the learned model and the calibration table is smaller than the original size of an entire provenance graph structure, achieving data compression for storage. Based on the size of the graph structure, we can tune the hyperparameters of XGBoost max\_depth (the depth of the tree) and n\_estimators (the number of trees) to get the right model size and training time. Different sizes of provenance are used in our experiments, where the hyperparameter settings can be used as a reference in Section \hyperref[graphsize]{\uppercase\expandafter{\romannumeral3}-C-5}. 
Compared with the original graph size, the size of graph structure is very small. For the provenance graph of 1GB size, the size of the graph structure is usually less than 100 MB. Therefore, It is not difficult for LESS to train the model in storage. Since the complete provenance graph is used for training and all the mispredictions are corrected, the storage system saves information in a lossless fashion. Furthermore, the trained model is naturally obfuscated, leading to preventing log tampering and maintaining system security.

{\bfseries Structure Storage Design.}
The storage of graph structures involves three phases: graph structure vectorization, model training, and calibration. During the vectorization, we first transform the raw data into vector representations and employ XGBoost model to learn them. Finally we calibrate the mispredicted outputs. The trained model and calibration table are the final two files we need to store.

{\bfseries Graph Structure Vectorization.} 
Graph Structure Vectorization refers to the process that we initially transform the graph structure into vector representation suitable for model training. In practice, we first convert the graph structure into an adjacency list representation, where each node element corresponds to a node identifier, and the list of nodes contains IDs for each successor node of that node. Subsequently, we compute Delta coding on the neighboring entries of the adjacency list. Since the IDs of nodes in the provenance graph are typically sequentially numbered, the data usually share long prefixes, leading to high redundancy. By storing the difference of the latter value minus the former value, we can make a significant reduction in the length of the values. For example, the list $25534\to[25535, 25536, 25537, 25542, 25558]$ is processed as $25534\to[1, 1, 1, 5, 16]$. Furthermore, we calculate the number of adjacent identical Delta coding values and finally transform the original adjacency list into a 2-tuple list, i.e., $25534\to[(3,1),(1,5),(1,16)]$, further reducing the size of data storage.

Subsequently, we further flatten the 2-tuple list into a one-dimensional vector by connecting them head-to-tail, which is suitable for the classification task. The 2-tuple lists are represented by 13 values, ranging from 0 to 12. We convert all the data of the list elements into the single-digit number, [0,9], with the value 10 serving as the delimiter between numbers of the original adjacency lists, and values 11 and 12 used as delimiters to separate each list and indicate the end of the converted lists respectively (e.g., converting \verb|"|25534\verb|"| into the sequence \verb|"|2, 10, 5, 10, 5, 10, 3, 10, 4, 10\verb|"|). In a classification task, the more classes we have, the more difficult it is for the model to make correct prediction. To mitigate this challenge and improve classification accuracy, we minimize the number of classes by using the vectors with element values ranging from 0 to 12, to represent the graph structures. 


{\bfseries Model Training. } 
We use the XGBoost model to learn graph structure vectors. The XGBoost standard library\cite{xgboostgithub} is utilized for training, with the graph structure vectors as inputs and the predicted subsequent characters as outputs. We terminate training when the model's accuracy stabilizes. Finally, we save the trained XGBoost model. All the practices adhere to standard machine learning practices. When utilizing the trained model to rebuild the graph structure, we input a fixed sequence as a starting point for prediction, thus obtaining a predicted vector of equal length to the input vector. 

{\bfseries Calibration. }
Since the trained model's accuracy does not reach 100\%, meaning there are discrepancies between the model's output and the original data, it is necessary to construct a calibration table to rectify these errors. Additionally, as the prediction outputs have the same length as the graph structure vectors, the calibration table only needs to record both the positions of incorrect characters and their corresponding accurate values. As a result, through the utilization of this calibration table, we are able to correct mispredictions and return the original graph structure vectors. In the calibration table, we use  ($\delta$, $v$) to represent the positions of predicted errors and the correct values, where $\delta$ is the offset position of mispredictions, $v$ means the corresponding correct character. As shown in the Figure \ref{LESSoverview}, the (3,10) in the calibration table indicates that the correct character $'10'$ is offset by 3 from the position 0, and the (3,1) indicates that the correct character of $'1'$ is offset by 3 from the previous calibrated position.

We establish a set of encoding rules to reduce the storage cost of the calibration table. Specifically, we allocate one byte to store the values of $v$, which range from 0 to 12. There are two conditions for $\delta$: if its value falls within the range of 0 to 127, we use one byte to store it and the first bit position is 0 as the flag bit; if the $\delta$ value is not in this range, we use two additional bytes to store it and the first bit position is 1 as the flag bit. We discuss the probability that the storage of $\delta$ value exceeds 1 byte and 3 bytes. The proposition \verb|"|storing $\delta$ exceeds 1 byte\verb|"| is equivalent to the proposition \verb|"|continuous correct prediction for $2^7=127$ bits or more\verb|"|. Taking the accuracy of 0.9 as an example, this probability is $< 1.5e-6$, while the probability of the $\delta$ value length exceeding 3 bytes is $< 0.9^{(2^{24}-1)}$.

In conclusion, during the graph structure storage phase, LESS converts the input provenance graph structure into a model and a calibration table, achieving lossless compression. The vector data outputted by the model is corrected with the calibration table to retrieve the original vector which represents the graph structure. Following decoding (the inverse process of the graph structure vectorization), the original provenance graph structure can be reconstructed. In our experiments, the sizes of the model and calibration table are typically less than 2\% of the original graph structure size.

\subsection{Graph Attribution Storage}\label{Graph Attribute Storage}

{\bfseries Locality of Log. }
We define the locality of logs as the temporal and spatial similarities of logs, which forms the basis for compressing provenance graph attributes. Logs from the same host typically exhibit a high degree of similarity in their entries within a specific time range, which may stem from various reasons. From a micro perspective, the locality of logs is partly attributed to the Locality Principle, where the same instruction is executed multiple times or the same data is used repeatedly over extended time intervals. From the macro level, the server's business tends to remain stable over time, with different clients requesting the same service and generating logs with a large number of identical segments, differing only in some parameters of the fields. At the system and network level, a single operation usually invokes the same system call multiple times or sends numerous similar network packets, which results in minimal variation between neighboring lines of logs. The locality of logs on a provenance graph is manifested by interconnected nodes and edges that share high similarities in attributes or by the presence of a significant number of attributes of edges between two nodes.

\begin{figure*}[htbp]
\centering
\subfigure[Locality on DARPA TC]{
\includegraphics[width=0.315\textwidth]{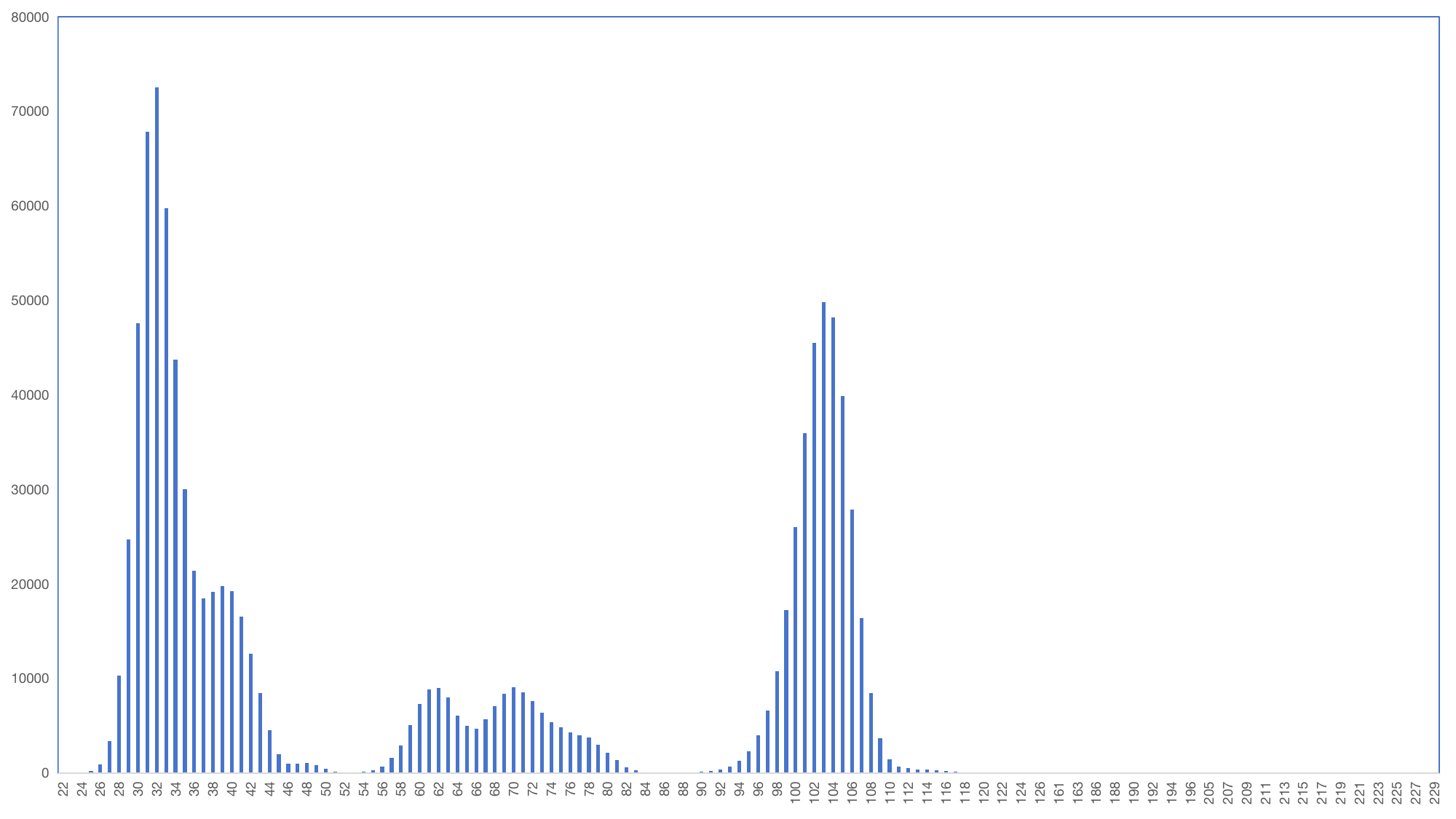}
}
\subfigure[Locality on DARPA OpTC]{
\includegraphics[width=0.315\textwidth]{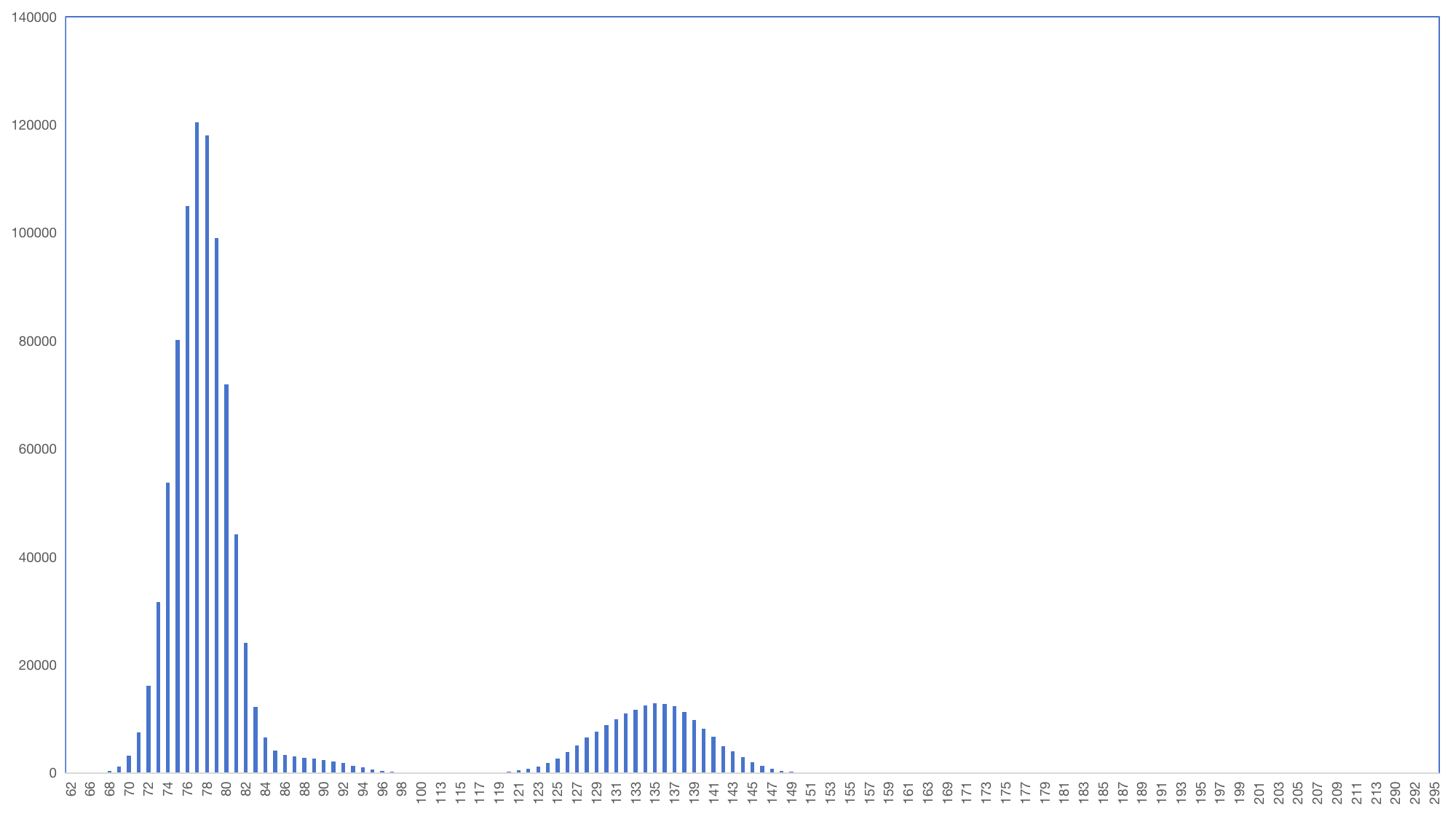}
}
\subfigure[Locality on Linux Audit]{
\includegraphics[width=0.315\textwidth]{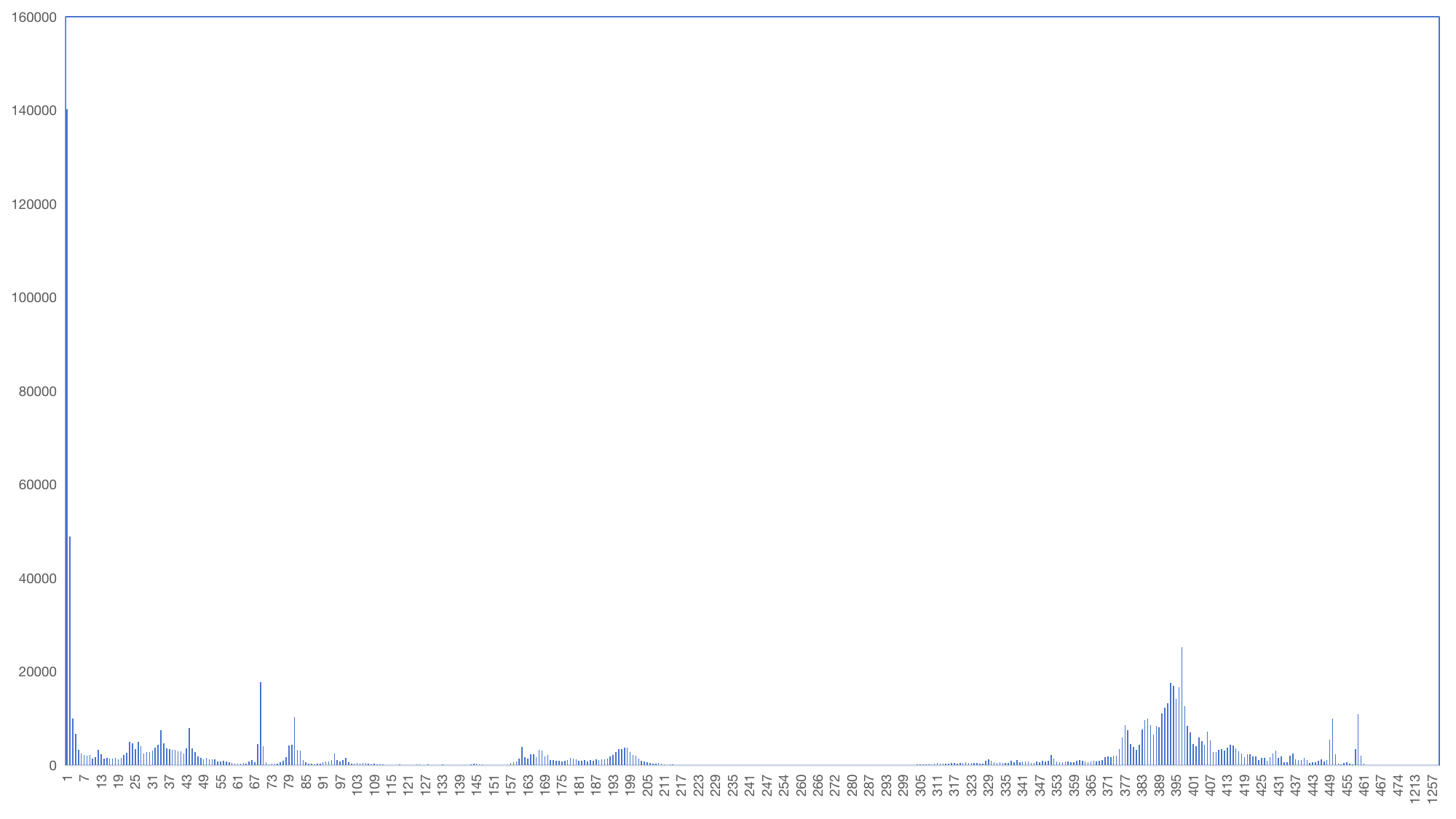}
}
\caption{Locality on different logs.}\label{Fig3}
\label{locality}
\end{figure*}

Edit Distance (Levenshtein Distance)\cite{editdistance} is a quantitative measure to calculate the similarity between two strings, which indicates the minimum number of editing operations, insertions, deletions, or substitutions, required to convert one string into another. We utilize Edit Distance to quantify the similarity between two log entries across three datasets: the DARPA TC, the DARPA OpTC, and a self-collected Linux Audit, each containing 100,000 rows of logs. We use a window size of 10 and a step of 1 to compute the Edit Distance between each log and the subsequent 10 logs. Then, we sum up the number of each Edit Distance value, as results depicted in Figure \ref{locality}. The percentage of Edit Distances less than 50 in DARPA TC exceeds 50\%, while the percentage of Edit Distances less than 100 in DARPA OpTC surpasses 80\%. Additionally, the percentage of Edit Distances less than 100 in the Linux Audit log is over 40\%. These statistics demonstrate varying levels of log locality across different logs.

The locality of logs can be utilized for log compression. By extracting the identical parts of the log entries and recording them once, we can reduce the log redundancy to achieve lossless compression. To achieve the theoretical maximum compression ratio, the similarities among all log entries need to be calculated. However, due to the high computational complexity, this method can be time-consuming. In LESS, we develop an optimized approach for similarity calculation that achieves lower computational complexity. Based on the similarity between log entries, we are able to organize logs into a tree data structure, with parent nodes representing common parts and child nodes denoting differences from the parent nodes. As a result, we construct a minimal attribute tree containing all attributes of the provenance graph.

{\bfseries Attribution Storage Design. }
The graph attribute storage is divided into similarity matrix computation and attribute tree generation. We compute the similarity matrix to measure the differences between attribute strings. According to the similarity matrix, we select the similar attribute strings and merge them into a tree. The resulting attribute tree occupies significantly less space than the original set of attribute strings, thus realizing the compressed storage of graph attributes.

{\bfseries Similarity Matrix Computing. }
To facilitate the calculation of similarity between attribute entries, we first encode the attribute strings into vector representations using the Bag-of-words model\cite{bagofword}, and then compute the Manhattan distance between the vectors as the distance between the attribute strings. 

The Bag-of-words model is a way to represent texts as a multiset. In this model, the order and correlation between words are ignored. It assumes that each word's occurrence in the text is independent of others. A character-level Bag-of-words model can be used to encode an attribute string of any format into a fixed-length vector whose dimension is equal to the number of unrepeated characters in the dataset.

We assume that the set of unduplicated character in the dataset is D. 
$$D=\{c_1,c_2,...,c_n\}$$

For any string $S$ composed of characters in $D$, define mapping $F$,  $F(S)$ is a character-level Bag-of-words representation of $S$, $Z^{|D|}$ is an integer space with cardinality of $D$ as dimension.
$$F:S\to Z^{|D|}$$

We define $I$ as a indicator function, and the i-th dimension of the generated vector can be represented by the following formula:
$$F(S)^{(i)}=\sum_{j=1}^{|D|}I(S_j=D_i)$$
In Figure \ref{LESSoverview}, for the attribute string $n1 = \verb|"|aaabac\verb|"|, n2 = \verb|"|ababac\verb|"|, n3 = \verb|"|baaaba\verb|"|$, we set the character set $D = \{a, b, c\}$.
Then the vectors of $n1$, $n2$, and $n3$ are $v1 = (4,1,1), v2 = (3,2,1), v3 = (4,2,0)$.

Once we convert attribute strings into attribute vectors, we can efficiently measure the similarity between attributes. We set a fixed-size window, scan through the entire attribute list, and compute the Manhattan distance between all attributes within the window. We suppose $p$, $q$ are two vectors of length $n$, and the Manhattan distance is defined as:
$$d(p, q)=\sum_{i=1}^n|p_i-q_i|$$

For multiple given attribute strings, we calculate the similarity matrix by window scanning method. For a total of $N$ attribute strings, each attribute string takes subsequent $M-1$ strings attribute strings to calculate Manhattan distance. The algorithm calculation process is presented as Algorithm \ref{algorithm1}. 
For example, 2 in the matrix is calculated from $v1$ and $v2$, and represents the Manhattan distance between $v1$ and $v2$ in Figure \ref{LESSoverview}.
\begin{algorithm}
\SetAlgoLined
\LinesNumbered
\KwIn{Attribute List $S$, Window Size $M$}
\KwOut{Similarity Matrix $D$}
N = $\mid$S$\mid$ \;
E = \{ \}\;
\For{i in \{1,...,N\}}{
E[i] = Bag\_of\_words\_Encode(S[i])\;
}
\For{i in \{1,...,N-1\}}{
\For{j in \{1,...,M-1\}}{
\eIf{i + j $\leq$ N}{
    D[i,j] = Manhattan\_Distance(E[i], E[i+j])\;
}{
    D[i,j] = +$\infty$\;
}
}
}
\Return D
\caption{Similarity Matrix Computing.}
\label{algorithm1}
\end{algorithm}


The combination of the Bag-of-words model with the Manhattan distance is the number of different characters between the statistical attribute strings, which is an approximate estimate of the Edit Distance\cite{editdistance}. Assuming that there are two strings of length $n$ and $m$, the time complexity of the Edit Distance is $O(nm)$, and the time complexity of calculating the Manhattan distance using the bag-of-words model is only $O(n+m+k)$ ($k$ is the character set size), we compare the two methods in Section \hyperref[attrivutevectorizationsec]{\uppercase\expandafter{\romannumeral3}-C-2}. The optimized method combining the Bag-of-words with Manhattan distance resulted in a 3.8\% increase in attribute disk usage, while reducing time expenditure by 31.9\% compared to direct computation of Edit Distance. In particular, the combination method of the Bag-of-words model with the Manhattan distance does not introduce any error into the attribute tree generation. The similarity matrix will only help us select similar attributes, and will not affect the final stored content. In the case of \verb|"|firefox, process, 89\verb|"| and \verb|"|firefox, process, 98\verb|"|, although they get a very high degree of similarity according to our method, we will handle them correctly in the next step.


{\bfseries Attribute Tree Generation. }
The similarity matrix can be regarded as an expression of an undirected graph, on which we can compute a minimum spanning tree (MST). Upon obtaining the MST, we further calculate edit operations between attribute strings. These edit operations are stored in the MST to form a minimal attribute tree, achieving compression of provenance graph attributes. The total distance between attribute vectors on the MST is almost minimized, implying that by recording the fewest edit operations, we can reduce the space occupied by attribute storage to the utmost. During MST computation, we set a parameter named maximum distance. Attribute vector pairs with distances exceeding this maximum threshold are not included in the same tree, as the space required to store edit operations may exceed the space that is needed for directly storing attribute strings.

We define the format of the attribute tree nodes, which are represented by triples: \textit{$E$=(Parent node ID, current node ID, edit operation set)}.
An edit operation set consists of multiple edit operations. The edit operation set and the edit operation have the following definitions:  \textit{Edit operation set =(Edit operation 1, edit operation 2,... Edit operation $n$)}.
There are three types of edit operations:
\textit{Insertion}=(\verb|"|insert\verb|"|, insert\_pos, insert\_string), 
\textit{Deletion}=(\verb|"|delete\verb|"|, start\_pos, stop\_pos), 
\textit{Substitution}=(\verb|"|substitute\verb|"|, substitute\_pos, substitute\_string).
\textit{Insertion} needs to specify the insertion position and the content of the string to be inserted. \textit{Deletion} specifies the deletion start, and end position. \textit{Substitution} specifies the replacement start position and the content of the string to be replaced.

The algorithm for building the attribute tree is shown in Algorithm \ref{algorithm2}. Firstly, the minimum spanning tree is calculated on the similarity matrix to obtain the set of edges that need to generate edit operations. When the distance of edges is less than the max distance, the corresponding edit operations are generated. Finally, for the nodes that are not added to the tree, we directly record the attribute values of the nodes. For the attribute strings $n1 = \verb|"|aaabac\verb|"|, n2 = \verb|"|ababac\verb|"|, n3 = \verb|"|baaaba\verb|"|$, the corresponding attribute tree of $n1$, $n2$, and $n3$ is [(1, 2, (\verb|"|replace\verb|"|, 1, \verb|"|b\verb|"|))), (1, 3, (((\verb|"|delete\verb|"|,0,0),(\verb|"|insert\verb|"|,6,\verb|"|c\verb|"|)))]. The meaning of this attribute tree is that the first node ID is 2, whose parent is 1,  with an edit operation: replacing the character in position 1 (indexed from 0)  with \verb|"|b\verb|"|. The second node ID is 3, whose parent is 1, with an edit operation: deleting the initial character at index 0 and appending \verb|"|c\verb|"| at the end.

\begin{algorithm}
\SetAlgoLined
\LinesNumbered
\KwIn{Attribut List $S$, Similarity Matrix $D$, Max Distance max\_dis}
\KwOut{Minimum Attribute Tree T\_prop}
T\_min = minimum\_spanning\_tree(D)\;
T\_prop = []\;
visit = \{\}\;
\For{edge in T\_min}{
\If{edge.distance <= max\_dis}{ 
    \tcp{Get edit operation based on attribute strings.} 
    edit\_ops = getEditOps(S[edge.parent], S[edge.current\_id])\;
    \tcp{Record edit operation.}
    T\_prop.add((edge.parent, edge.current\_id, edit\_ops))\;
    visit.add(edge.current\_id)\;
}
}
\For{node\_id not in visit}{
    \tcp{Store nodes that are not visited.}
    T\_prop.add((node\_id, node\_id, ("insert", 0, S[node\_id])))\;
}
\Return T\_prop
\caption{Attribute Tree Generation.}
\label{algorithm2}
\end{algorithm}

In conclusion, in graph attribute storage, we store frequently occurring substrings in attribute strings only once, and store edit operations for the different parts relative to common substrings. This approach minimizes the storage of redundant information to the greatest extent. Furthermore, within the attribute tree, all attribute strings can be fully restored by executing edit operations downward from the root node. Thus, we achieve lossless compression tailored specifically for graph attributes.

\subsection{Query}\label{Query}

\begin{figure}[htbp]
  \centering
  \includegraphics[width=\linewidth, trim=6mm 2mm 305mm 30mm, clip]{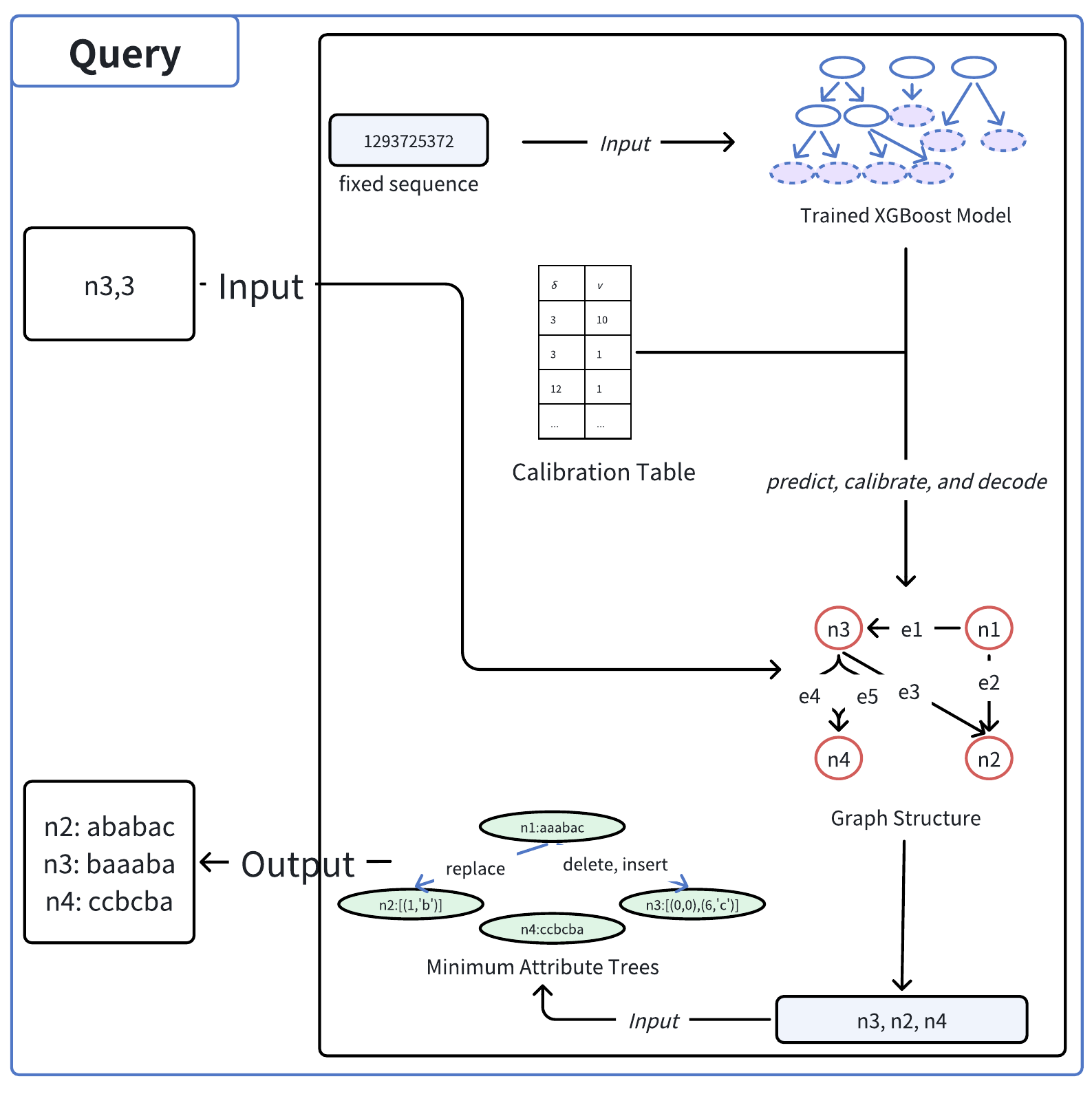}
  \caption{Overview of the querying process of LESS: We input (n3, 3) to query detailed information about node n3 and its subsequent 2 nodes, along with the related edges. In a cold query, LESS first restores the graph structure, obtain the IDs of targeted nodes and edges in the graph structure, then inputs the IDs to the minimum attribute tree and retrieves detailed information about all nodes and edges. (For simplicity, the edge querying process is not illustrated here; this process follows the same procedure as the above node querying process.)}
  \label{queryofless}
\end{figure}

LESS supports back-tracing and forward-tracing queries on the provenance graph. Given the node IDs and the threshold (the sum of the number of nodes and edges included), LESS can return detailed information about the nodes and connected edges. When querying, LESS first rebuilds the entire provenance graph structure, searches the IDs of the given and targeted query nodes in the graph structures, and then recovers the complete attributes of these nodes and edges, as shown in Figure \ref{queryofless}.

When recovering the graph structure, we first use a fixed sequence of pre-preserved in the calibration table as the starting point for XGBoost model to predict results. Then we obtain the complete predicted vectors and correct the errors by calibration tables, obtaining the correct vector data. Subsequently, we carry out a decoding (inverse process of the graph structure vectorization) to rebuild the final graph structure. A search is conducted on the obtained graph structure to get the ID lists of relevant nodes and edges. After inputting the ID list into the attribute query module, we first search for the tree nodes where the target ID is located and then recursively search upward for the node's parent node until we reach the root node. We read the attribute string stored in the root node and then apply edit operations along the search path until we reach the targeted node from the root. Finally, we receive a complete targeted attribute string. By repeating this process, we can recover the attributes of all the nodes and edges in the ID list.

\section{Evaluation}

\begin{table*}[hbtp]
    \centering
  \caption{Overview of Datasets}
  \label{tab:datasets}
\begin{tabular}{ccccccccc}
\hline
Name & Dataset    & Log Size  & Graph Size & Edges      & Vertices   & Structure Size & Attribute Size & File Name                            \\ \hline
D1   & DARPA TC   & 6.82GB    & 1.63GB     & 9,095,615  & 1,182,293  & 67MB           & 621MB          & ta1-trace-1-e5-official-1.bin.101.gz \\
D2   & DARPA TC   & 6.71GB    & 1.65GB     & 9,085,593  & 1,187,728  & 65MB           & 613MB          & ta1-trace-2-e5-official-1.bin.16.gz  \\
D3   & DARPA TC   & 6.81GB    & 1.63GB     & 9,091,104  & 1,200,821  & 66MB           & 615MB          & ta1-trace-1-e5-official-1.bin.50.gz  \\
D4   & DARPA TC   & 6.80GB    & 1.64GB     & 9,101,879  & 1,200,780  & 63MB           & 610MB          & ta1-trace-3-e5-official-1.bin.gz     \\
D5   & DARPA TC   & 6.81GB    & 1.62GB     & 9,132,726  & 1,114,267  & 58MB           & 609MB          & ta1-trace-3-e5-official-1.bin.26.gz  \\
S1   & DARPA OpTC & 1.20GB    & 0.7GB      & 2,162,546  & 2,082,587  & 18MB           & 464MB          & AIA-101-125                          \\
S2   & DARPA OpTC & 2.45GB    & 1.4GB      & 4,417,515  & 4,246,718  & 37MB           & 949MB          & S1 + AIA-151-175                     \\
S3   & DARPA OpTC & 3.53GB    & 2.1GB      & 6,385,169  & 6,106,283  & 54MB           & 1372MB         & S2 + AIA-201-225                     \\
S4   & DARPA OpTC & 4.63GB    & 2.7GB      & 8,347,544  & 8,028,468  & 71MB           & 1791MB         & S3 + AIA-301-325                     \\
S5   & DARPA OpTC & 5.52GB    & 3.2GB      & 9,948,743  & 9,550,801  & 85MB           & 2134MB         & S4 + AIA-51-75                       \\
A*   & -          & 208.00 GB & 22.07GB    & 85,193,120 & 14,156,636 & 5.7GB          & 5.8GB          & -                                    \\ \hline
\end{tabular}
\end{table*}

\subsection{Experiment Setup}

LESS is implemented in Python 3.9.19 using sklearn 1.4.2, numpy 1.26.4, and pandas 2.2.2. All experiments are done on a Ubuntu 18.04 machine equipped with 256GB of RAM, 2 $\times$ Intel(R) Xeon(R) Gold 5218R cpus @ 2.10GHz, and 8 $\times$ NVIDIA Corporation GA102 [GeForce RTX 3090] GPUs.

{\bfseries Datasets. } Our experiments consist of 11 datasets collected by the trace group in DARPA transparent computing (DARPA TC) engagement $\#5$\cite{darpatc}, DARPA operationally transparent cyber (Darpa OpTC)\cite{darpaoptc} and ourselves. As detailed shown in Table \ref{tab:datasets},  columns 3 and 4 specify the size of the log files and the provenance graph processed by the logs; columns 5 and 6 are the number of edges and nodes in the provenance graph; while the columns 5 and 6 indicate the size of the preprocessed provenance graph structures and attributes that are inputs to the LESS. The last column is the name of the experiment file in the public link \cite{Tclink, Optclink}. The OpTC file is selected from ecar-bro/benign/20-23Sep19. We combine the files in OpTC to form datasets of increasing size for ablation study. Dataset Sn is combined from a new file and the dataset Sn-1. In addition to the public dataset, we collect a large dataset from our own simulated enterprise server. We deploy a Web server running Apache2, Nginx, and more than 40 websites. We run 4 clients, each of which uses Python scripts to randomly access the Web service. We launch the Audit framework on the server to collect system logs. We select one of these files and construct the provenance graph with Spade\cite{spade}, which is the A* in the table.


In the graph structure storage, the settings for XGBoost model are the max\_depth 3, n\_estimaters 3 and the loss function set as multi-class classification Softmax. The window size is set at 4 for calculating the similarity matrix in the graph attribute storage, and the max distance between neighboring nodes for the attribute tree generation is set at 60.

\subsection{Storing \& Querying}

We compare LESS, LEONARD\cite{leonard}, and SEAL\cite{seal} in terms of disk usage, storage time costs, query speed, and memory usage. All experiments are carried out on five datasets (from D1 to D5). LEONARD is implemented and set on its publicly available datasets\cite{leonardcode}. We reproduce SEAL based on the published paper\cite{seal}. We only present the key data in the paper due to limited space, the full data is available in our open-source link.


\begin{table*}[hbtp]
    \centering
  \caption{Performance of LESS}
  \label{tab:LESS Performance}
\begin{tabular}{cccccccccccccccc}
\hline
Dataset & Model & Cal. T & Tree & Disk & Prep. & A.1   & A.2   & A.3   & B.1    & B.2    & Time & Warm\_up & Query & Ave. Mem  & Max Mem   \\ \hline
D1      & 65KB  & 347KB  & 37MB & 38MB & 73.0s & 76.2s & 15.5s & 8.7s  & 83.2s  & 665.8s & 922.4s     & 95.7s    & 14,281      & 2,281.9MB & 5,858.9MB \\
D2      & 63KB  & 292KB  & 37MB & 38MB & 77.7s & 76.8s & 12.7s & 8.2s  & 82.9s  & 655.3s & 913.6s     & 89.5s    & 14,923      & 2,533.9MB & 7,091.3MB \\
D3      & 66KB  & 417KB  & 38MB & 39MB & 82.7s & 81.9s & 16.1s & 18.7s & 115.4s & 724.4s & 1039.2s    & 91.0s    & 15,274      & 2,278.4MB & 5,842.9MB \\
D4      & 62KB  & 235KB  & 33MB & 34MB & 79.6s & 77.4s & 12.3s & 6.7s  & 84.0s  & 644.3s & 904.3s     & 72.0s    & 15,048      & 2,267.0MB & 5,815.7MB \\
D5      & 63KB  & 183KB  & 33MB & 33MB & 76.2s & 77.9s & 8.1s  & 6.3s  & 81.1s  & 624.7s & 874.3s     & 68.1s    & 15,092      & 2,479.4MB & 7,002.8MB \\ \hline
\end{tabular}
\end{table*}

\begin{table}[hbtp]
  \caption{Performance of LEONARD}
  \label{tab:LEONARD Performance}
\begin{tabular}{cccccc}
\hline
Dataset & Disk Usage & Time Costs & Query Speed & Ave. Mem   & Max Mem    \\ \hline
D1      & 194MB      & 5,356.2s   & 892         & 20,883.7MB & 35,329.9MB \\
D2      & 194MB      & 5,664.3s   & 756         & 21,205.2MB & 36,203.3MB \\
D3      & 193MB      & 6,289.8s   & 761         & 20,569.7MB & 35,193.4MB \\
D4      & 184MB      & 5,979.5s   & 859         & 20,169.7MB & 34,107.0MB \\
D5      & 187MB      & 5,903.7s   & 806         & 20,111.4MB & 35,391.4MB \\ \hline
\end{tabular}
\end{table}

\begin{table}[hbtp]
  \caption{Performance of SEAL}
  \label{tab:SEAL Performance}
\begin{tabular}{cccccc}
\hline
Dataset & Disk Usage & Time Costs & Query Speed & Ave. Mem & Max Mem \\ \hline
D1      & 98MB              & 160.5s                   & 146,368                 & 4,691.2MB            & 8,123.1MB        \\
D2      & 97MB              & 157.4s                   & 149,105                 & 4,670.3MB            & 8,261.9MB        \\
D3      & 106MB             & 174.9s                   & 151,196                 & 4,762.9MB            & 8,274.7MB        \\
D4      & 96MB              & 162.6s                   & 155,441                 & 4,710.7MB            & 8,204.0MB        \\
D5      & 94MB              & 151.0s                   & 151,650                 & 4,717.5MB            & 7,886.3MB        \\ \hline
\end{tabular}
\end{table}
{\bfseries Disk Usage.}
We measure the total disk usage of LESS, LEONARD, and SEAL. The total disk usage of LESS (columns 5 \verb|"|Disk\verb|"| of the Table \ref{tab:LESS Performance}) includes the costs of the trained XGBoost model, calibration table, and attribute trees (including node attribute trees and edge attribute trees) as shown in the column 2-4 of the Table \ref{tab:LESS Performance}. For LEONARD, the total storage costs encompass the reference tables, indexes, LSTM models, and calibration tables.


From Table \ref{tab:LESS Performance}, \ref{tab:LEONARD Performance} and \ref{tab:SEAL Performance}, we can observe that LESS costs significantly less disk space to store the five datasets compared to LEONARD and SEAL, with an average space usage of only 19.1\% and 37.1\%.  Compared to the provenance graph without compression, LESS achieves a disk usage of only 2.03\%, 1.97\%, 2.04\%, 3.06\%, and 3.06\%, respectively, with an average of only 2.48\% space usage. LESS outperforms LEONARD in terms of disk usage, 5.23$\times$ the average compression rate of LEONARD. When we evaluate the disk usage of each component, the trained models of both LEONARD and LESS incur the lowest storage cost (only 65.6KB on average for LESS). The calibration table and attribute tree have average storage costs of 480.8KB and 38.5MB. Notably, We find that the attribute tree occupies the highest space, 97.8\% of the overall disk usage on average (43.58MB/44MB). The attribute tree takes up a significant amount of space due to our retention of sufficient raw data, enabling lossless compression and thereby enhancing the accuracy of subsequent intrusion detection and attack investigation tasks.

LEONARD trains the DNN model to learn all the attributes of an entire provenance graph. The huge number of mispredicted outputs in training lead to a significant storage cost of the calibration table, occupying 146MB of space on average. In contrast, LESS uses the training models to store only the provenance structure, and the XGBoost we utilized requires a much smaller space size while achieving higher compression ratios. SEAL uses encoding to compress fields with the same prefix and monotonically incrementing. On some datasets, this works, but expert analysis may be required to identify the presence of compressible fields.


{\bfseries Storage Time Costs. }\label{Storage Time Costs}
We evaluate the storage time costs of LESS, LEONARD, and SEAL. We define that the storage time starts from storing the provenance graph to the completion of generating the final files (the XGBoost model, calibration table, and attribute trees) for storage. The total time costs are illustrated in Table \ref{tab:LESS Performance}. The time cost of LESS includes 6 phases: preprocessing(Prep.), graph structure vectorization(A.1), model training(A.2), calibration(A.3), similarity matrix computing(B.1), and attribute tree generation(B.2).


Compared to LEONARD, we observe that LESS spends 17.2\%, 16.1\%, 16.5\%, 15.1\%, and 14.8\% of time to complete the storage tasks on the five datasets(from D1 to D5), with an average of 15.9\%. The average time costs of LESS on graph structure vectorization, model training, and calibration table generation are 78.0s, 12.9s, and 9.7s, respectively. LEONARD spends 293.2s and 5047.9s on model training and calibration, respectively. LESS spends less time on model training and calibration, enabling us to complete the storage tasks faster than LEONARD. The highest time cost of LESS is consumed by the attribute tree generation (662.9s on average). LESS has a 6.29$\times$ improvement in storage speed compared to LEONARD. SEAL compression is the fastest of the 3 methods because SEAL only encodes the fields that can be compressed line by line.

{\bfseries Memory Usage. }
We evaluate the memory usage of the 3 methods in storage. We record the memory usage of the process every 0.5 seconds, and get the average and maximum memory usage for the entire storage procedure (column \verb|"|Ave. Mem\verb|"| and \verb|"|Max Mem\verb|"| in Table \ref{tab:LESS Performance}). LESS has the least memory usage of 3 methods. LEONARD's average memory usage is 8.7 times that of LESS, and SEAL's average memory usage is 2 times. LSTM training of LEONARD takes lots of memory. SEAL performs compression by file scanning, which typically requires loading the entire provenance graph into memory. Due to the small size of the graph structure, LESS requires only a small amount of memory in the structure storage. In the attribute storage, LESS uses window scans to read files, limiting the size of what needs to be loaded into memory.
 
{\bfseries Query Speed. }
 We explore the LESS's performance in supporting back-tracking queries. In the query experiment, we first randomly select 100 nodes from the provenance graph, search their subsequent nodes, and return the details of descendent nodes and associated edges. To avoid excessive output, we stop the query when the total number of returned nodes and edges reaches 4096. Before the LESS query, we need to recover the graph structure (the time required is the column \verb|"|Warm\_up\verb|"| in Table \ref{tab:LESS Performance}). We define the query cost as the period from the start of search execution to the return of results. The queried edge and associated nodes represent an event in the system. The query speed is obtained by dividing the number of queried events by the query time cost (column \verb|"|Query\verb|"| in Table \ref{tab:LESS Performance}).


On the five datasets, the query speed of LESS is 18.3$\times$ faster than that of LEONARD on average. LEONARD needs the model to iteratively predict and correct errors on each query which takes a long time. LESS only needs to perform Warm\_up(\verb|"|predict, calibrate, and decode\verb|"| in the Figure \ref{queryofless}) once, after which no model prediction is required. SEAL performs best on queries because SEAL preserves the structure information from the original provenance graph and queries only need to decode line by line.

In addition to forward querying, we also studied backward querying. Backward querying introduces an additional step of converting the adjacency lists, which represent the graph structure, into reverse adjacency lists, while keeping all other steps the same as forward querying (Figure \ref{queryofless}). Our experiments showed that backward querying only takes 9\% more time compared to forward querying due to this extra operation.


In conclusion, LESS has advantages in disk usage, storage time costs, query speed, and memory usage compared to LEONARD. SEAL performs better in terms of storage time costs and query speed, while LESS performs better in terms of disk usage and memory usage. In addition, SEAL and LESS are compatible because the provenance graph output by SEAL can still be compressed and stored by LESS.
 
\subsection{Ablation Study}
In this study, we develop a novel framework to support provenance graph storage and queries. We conduct five ablation studies to understand how LESS performs under different settings. If not specified, the parameters are kept the same as the default settings described in the previous section.

\begin{table*}[hbtp]
    \centering
  \caption{Performance of LESS on A*}
  \label{tab:LESS Performance on A*}
\begin{tabular}{cccccccccccccccc}
\hline
Dataset & Model & Cal. T & Tree & Disk & Prep. & A.1   & A.2   & A.3   & B.1    & B.2    & Time & Warm\_up & Query & Ave. Mem  & Max Mem     \\ \hline
A*      & 0.12 & 77.6  & 554.2  & 632 & 1,910.2 & 635.1 & 778.3 & 421.9 & 1,292.38 & 8,154.92 & 13,192.8  & 4,996.7 & 2,263       & 28,663.4 & 142,337.9 \\
A*/2    & 0.12 & 39.45 & 273.43 & 313 & 744.0  & 276.6 & 371.5 & 199.3 & 600.43   & 3,819.82 & 6,011.65  & 1,915.2 & 4,076       & 14,777.7 & 71,303.06 \\
A*/4    & 0.12 & 20.19 & 137.69 & 158 & 365.7  & 150.6 & 192.1 & 99.2  & 308.13   & 1,928.07 & 3,043.8   & 806.0   & 6,532       & 7,999.58 & 36,158.55 \\
A*/10   & 0.12 & 7.90  & 55.98  & 64  & 142.2  & 57.4  & 74.5  & 38.3  & 125.85   & 781.51   & 1,219.76  & 224.9   & 9,644       & 3,899.21 & 14,757.50 \\
A*/20   & 0.12 & 4.08  & 29.8   & 34  & 71.4   & 30.72 & 30.5  & 19.7  & 63.61    & 396.54   & 612.47    & 129.3   & 11,507      & 2,303.58 & 7,451.12  \\ \hline
\end{tabular}
\end{table*}
\subsubsection{Large File}\label{largefile}
We explore the performance of LESS on a large file A* and slices of A*. We cut out slices of 1/2, 1/4, 1/10, and 1/20 sizes of the original size of A* for storage and query. During model training, the settings for XGBoost model are the max\_depth 6, n\_estimaters 6. The results are presented in Table \ref{tab:LESS Performance on A*}, where we omit the units $MB$ and $s$ for a complete presentation. It can be seen from the table that the disk usage and storage time costs vary in the same way as the file size. Query and memory usage on slices perform better than A*. This means that for memory-constrained servers, it is easier to store and query large files by splitting them into appropriately sized slices. Slicing large files requires keeping track of the relationship between the slices, which is not part of LESS's function. On A* and its slices, we achieve a compression rate of about 2.9\%. We use CPU to train our model, demonstrating that LESS can still achieve good storage performance on servers without GPUs.

\subsubsection{Attribute Vectorization}\label{attrivutevectorizationsec}

\begin{figure}[htbp]
\centering
\subfigure[Attribute Storage Disk Usage]{
\includegraphics[width=0.22\textwidth]{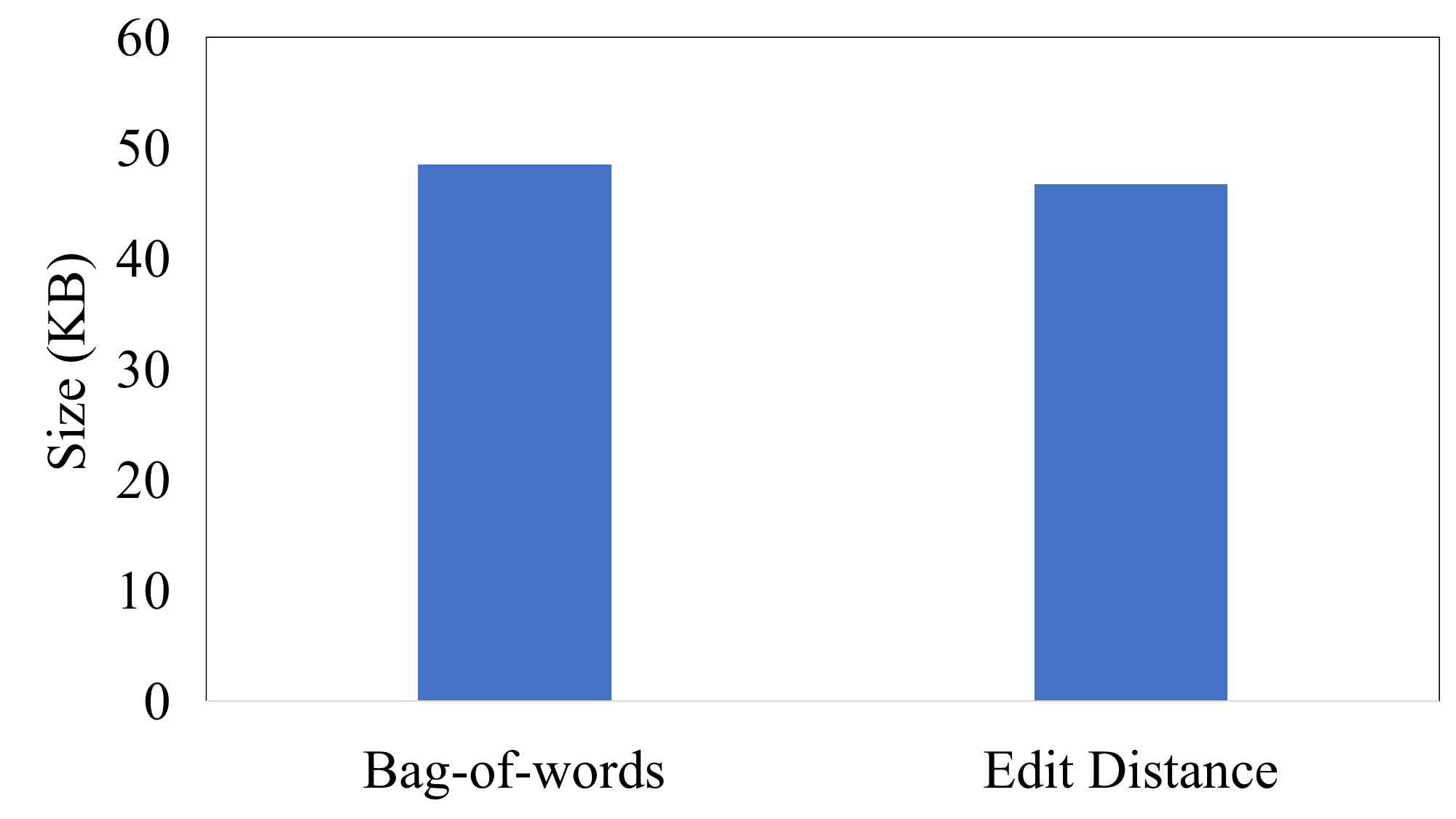}
}
\subfigure[Attribute Storage Time Costs]{
\includegraphics[width=0.22\textwidth]{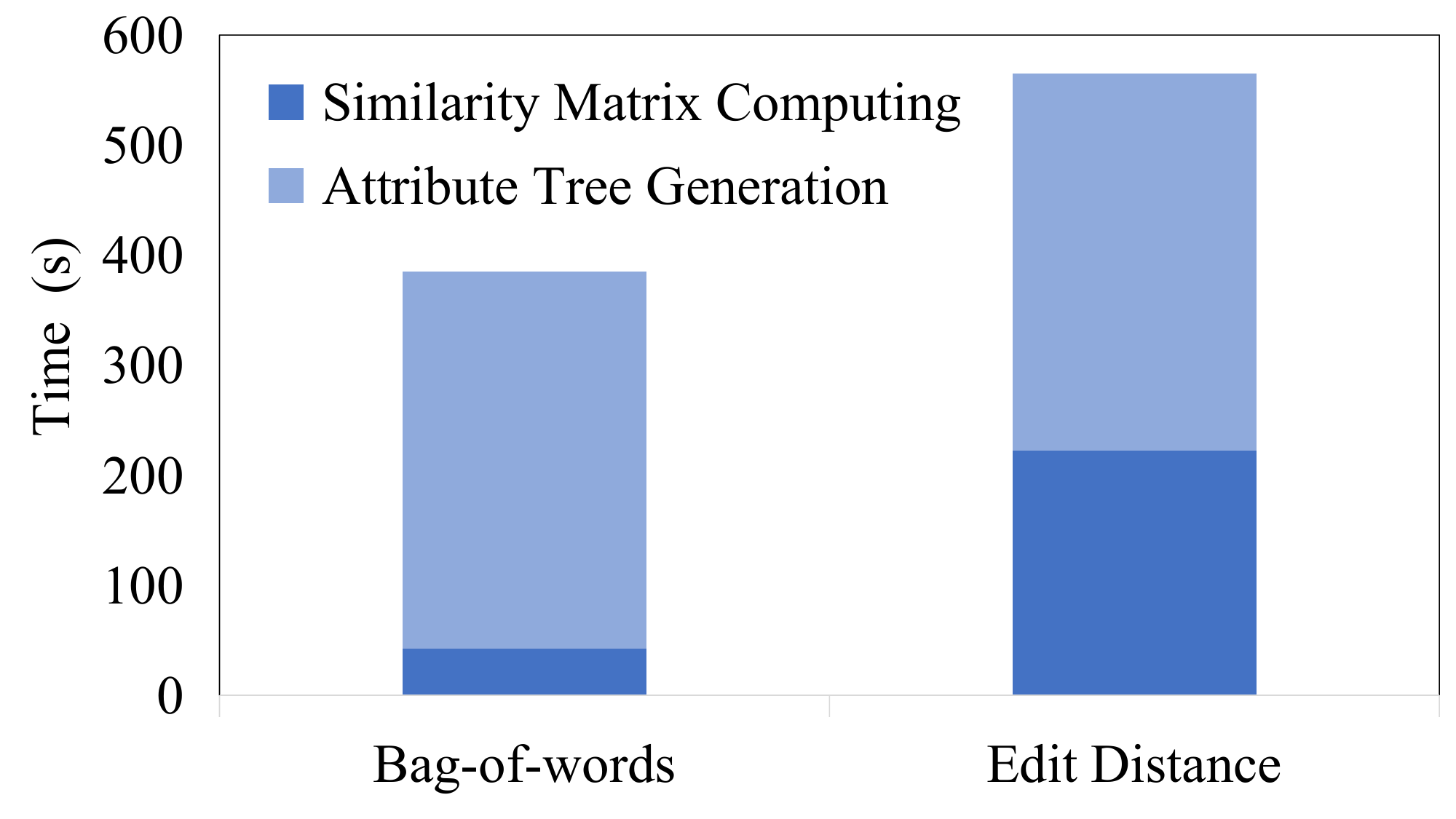}
}
\caption{Performance of LESS with the Bag-of-words and Manhattan distance and Edit Distance}
\label{attrivutevectorization}
\end{figure}


To compute the similarity matrix, we first use the Bag-of-words method to transform the graph attributes into vectors and use them to compute the Manhattan distance, achieving the similarity matrix. Directly calculating Edit Distances is of more accuracy, but costs more time. We also evaluate these two methods, recording the disk storage of the graph attributes and the time consumption of each phase. The results are shown in Figure \ref{attrivutevectorization}. The disk storage of the graph attribute storage for the similarity matrix computed by Bag of Words and Edit Distance is 48.5MB and 46.7M, respectively. In the similarity matrix computation phase, the two methods consume 42s and 222s, respectively. The time consumed by the two methods in the attribute tree generation is the same. The whole storage costs take 385s and 565s for Bag-of-words and Edit Distance.  After using Bag-of-words, the graph attribute storage increases by only 3.8\%, but the calculation speed of the similarity matrix is 5.29 times faster than that of using Edit Distance, with the graph attribute storage time overhead reduced by 31.9\%.

\subsubsection{Similarity Matrix Computing}

\begin{figure}[htbp]
\centering
\subfigure[Attribute Storage Disk Usage]{
\includegraphics[width=0.22\textwidth]{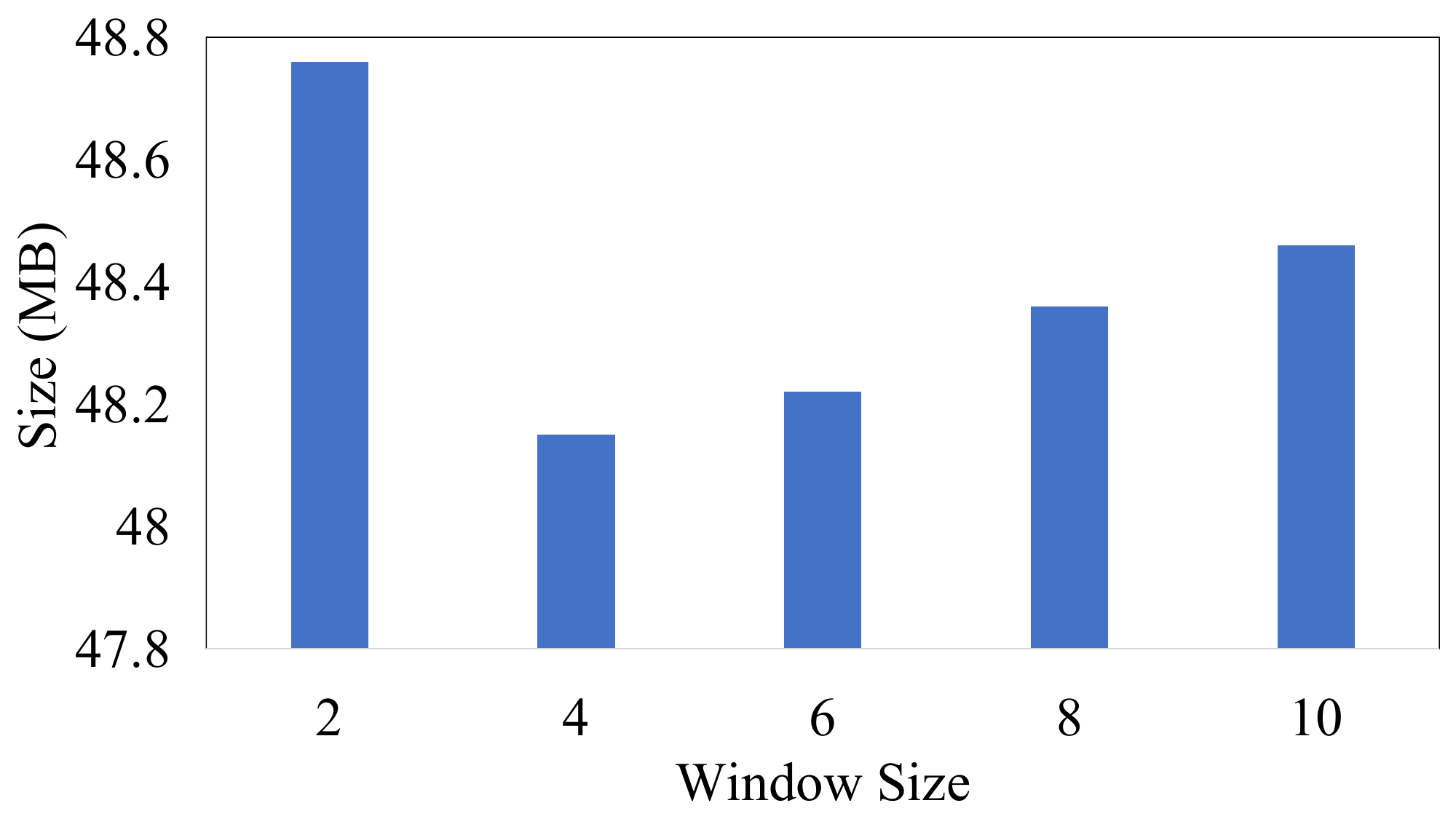}
}
\subfigure[Attribute Storage Time Costs]{
\includegraphics[width=0.22\textwidth]{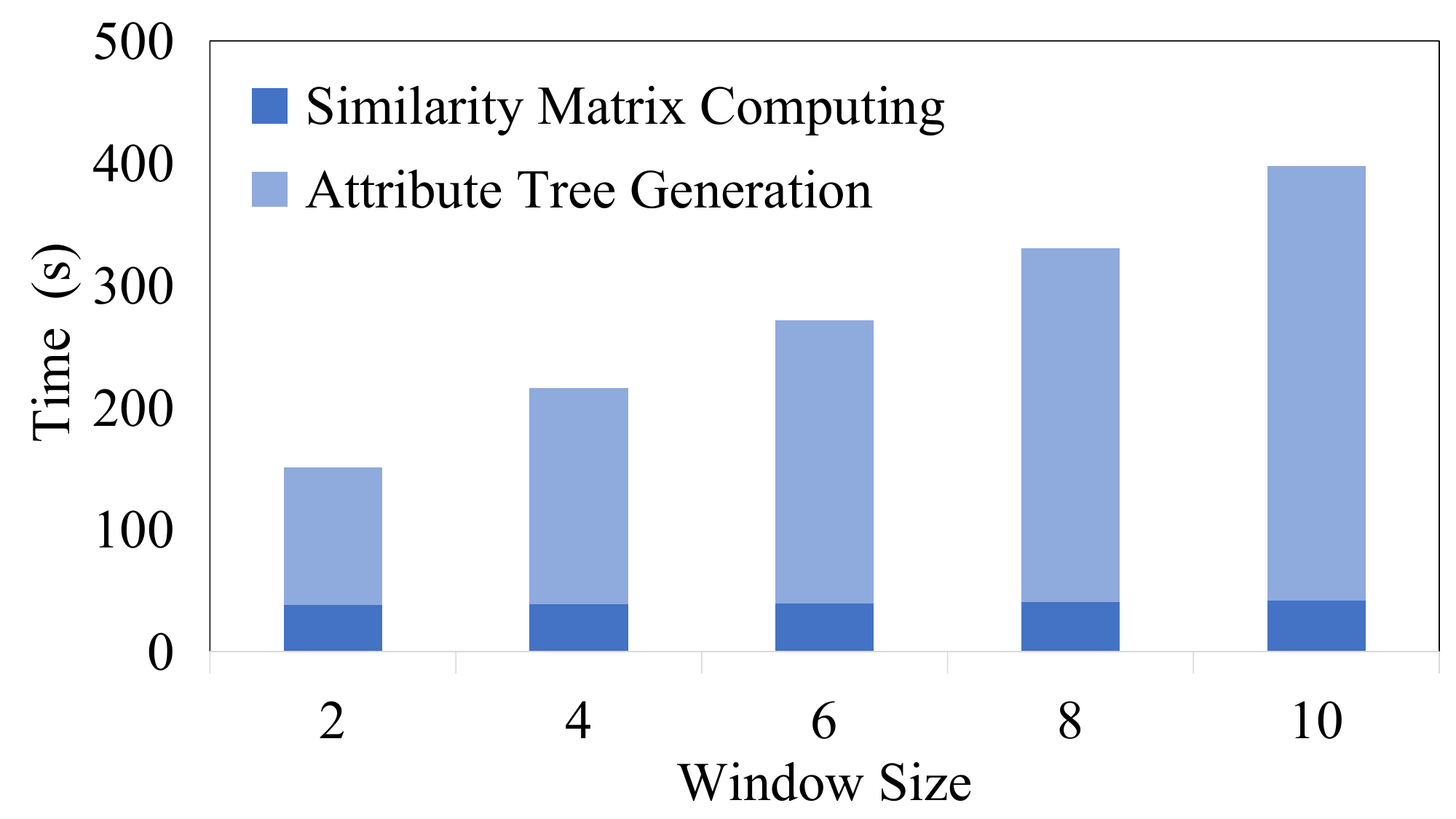}
}
\caption{Performance of LESS with different window size.}
\label{similaritymatrixcomputing}
\end{figure}


To measure the effect of different similarity matrix sizes on storage of attributes, we set the window scan widths from 2 to 10, generate similarity matrices with different widths, and record the attribute storage disk costs and attribute storage time costs under the window size settings. The results are shown in Figure \ref{similaritymatrixcomputing}. As the window width increases, the attribute disk cost first decreases and then increases. When the window width is too small, the similarity of two attribute strings that are not within a window is not utilized in the attribute tree generation; thus, the size of the attribute tree increases. When the window width is too large, since the value in our similarity matrix is an approximation of the Edit Distance, some attribute strings with high similarity may require very many edit operations. Recording these operations requires more space, which leads to a rise in attribute storage costs. As the window width increases, the time cost of attribute storage keeps increasing for all components. Relatively small window sizes can be used in LESS to achieve better compression ratios with lower time overhead. We also investigate other parameters including the max distance between nodes for attribute tree generation, which have no significant effect on the overall disk storage and time cost of LESS.

\subsubsection{Different Models}\label{differentmodelssec}

\begin{figure}[htbp]
\centering
\subfigure[Structure Storage Time Costs]{
\includegraphics[width=0.22\textwidth]{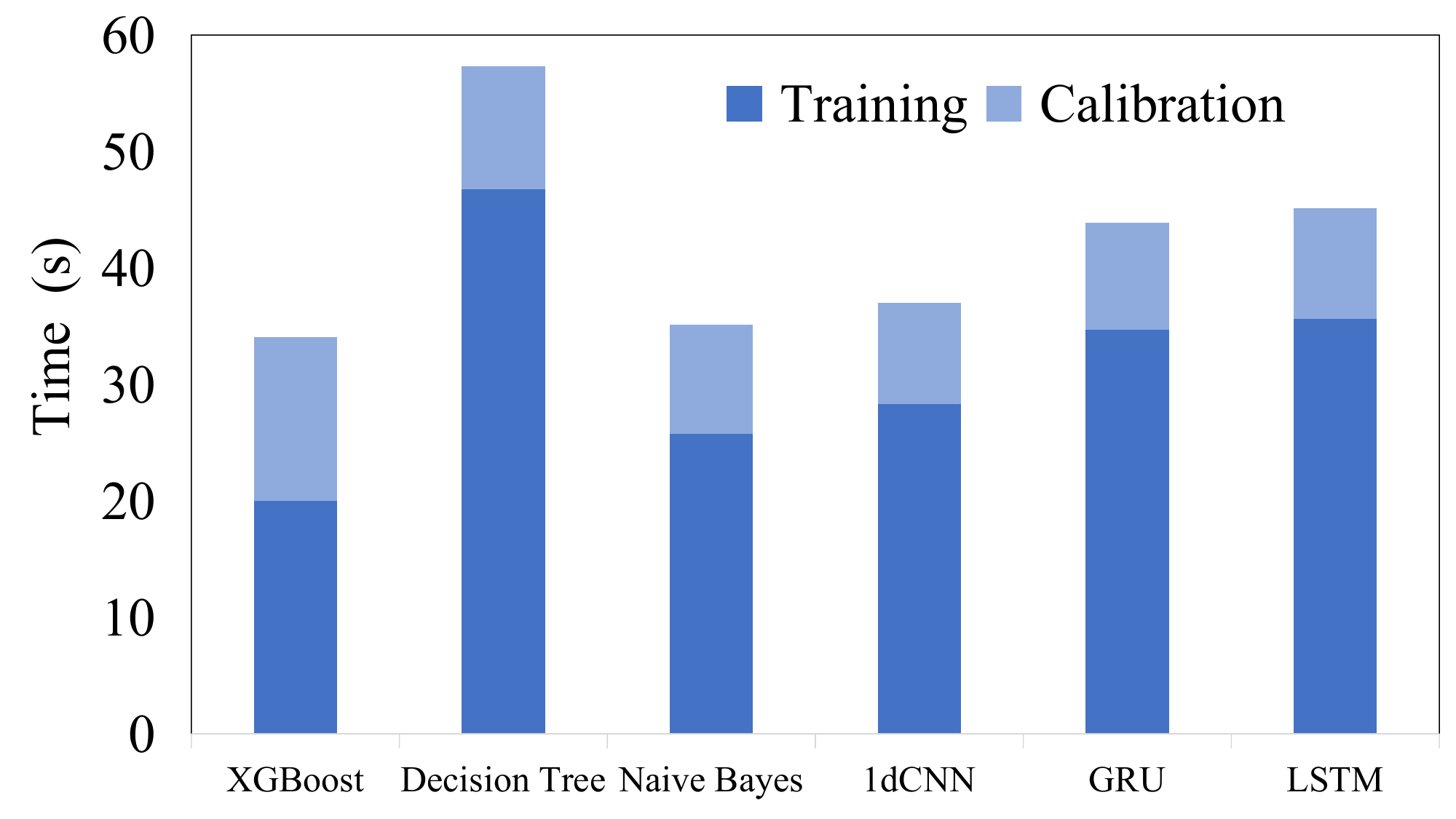}
}
\subfigure[Disk Usage and Accuracy]{
\includegraphics[width=0.22\textwidth]{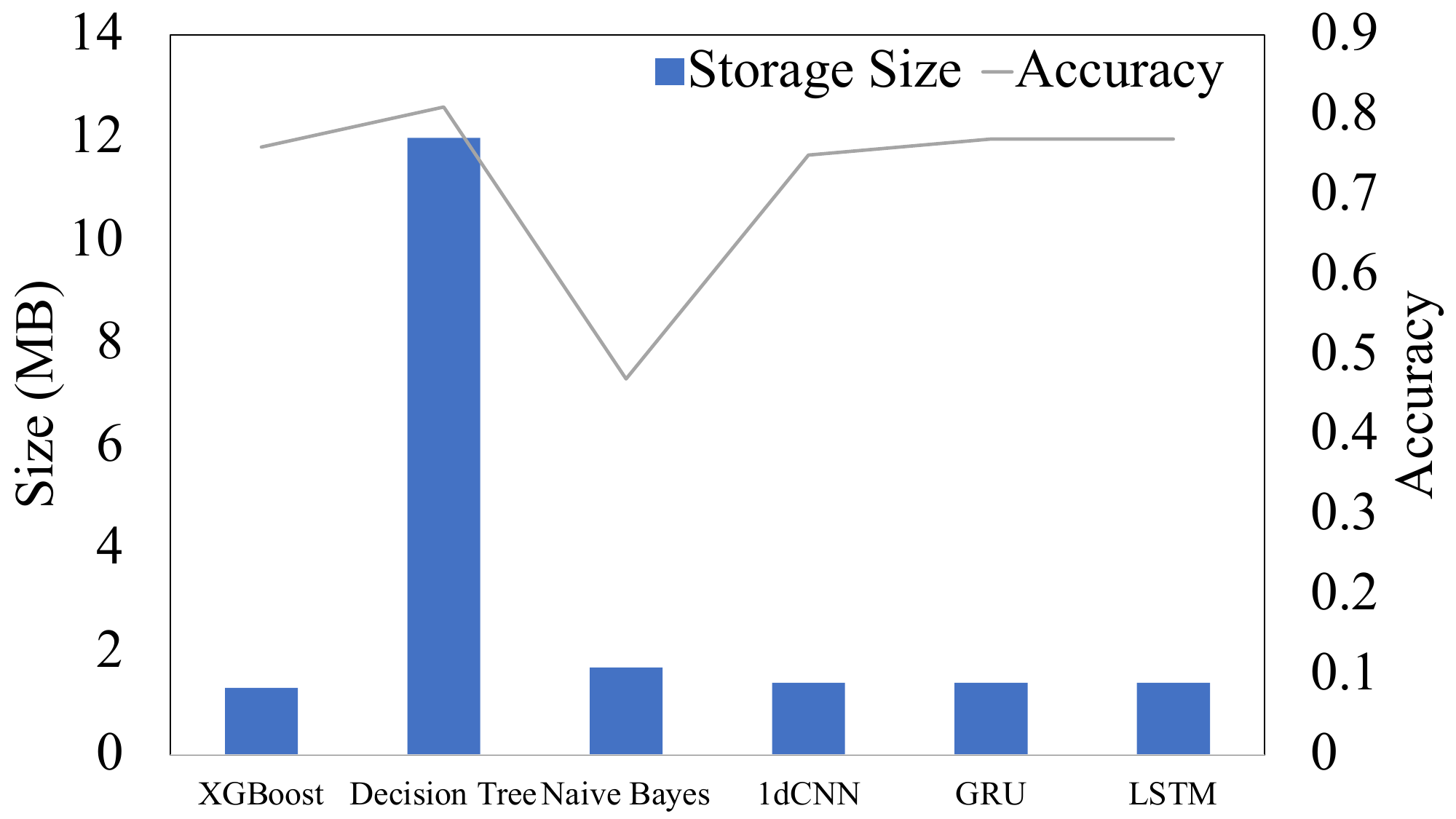}
}
\caption{Structure storage with different models}
\label{differentmodels}
\end{figure}

\begin{table*}[hbtp]
    \centering
  \caption{Performance of LESS on S1-S5}
  \label{tab:LESS Performance S1-S5}
\begin{tabular}{cccccccccccccc}
\hline
Dataset & Model & Calibration Table & Attribute Tree & Disk & Preprocess & A.1   & A.2   & A.3   & B.1    & B.2      & Time & Warm\_up & Query \\ \hline
S1      & 41KB  & 296KB             & 46MB           & 46MB       & 29.2s      & 17.8s & 2.6s  & 1.9s  & 36.4s  & 212.8s   & 300.7s     & 16.3s    & 13,912      \\
S2      & 41KB  & 625KB             & 95MB           & 96MB       & 58.8s      & 37.2s & 4.3s  & 4.2s  & 73.7s  & 478.1s   & 656.3s     & 39.9s    & 11,973      \\
S3      & 41KB  & 862KB             & 139MB          & 140MB      & 82.1s      & 59.5s & 7.1s  & 5.9s  & 107.6s & 720.8s   & 983.0s     & 58.6s    & 10,838      \\
S4      & 84KB  & 1.11MB            & 179MB          & 180MB      & 121.1s     & 76.2s & 14.4s & 8.3s  & 140.1s & 1,027.6s & 1,387.7s   & 69.6s    & 9,731      \\
S5      & 84KB  & 1.32MB            & 215MB          & 216MB      & 155.8s     & 90.7s & 17.6s & 11.4s & 171.4s & 1,362.2s & 1,809.1s   & 80.2s    & 8,591      \\ \hline
\end{tabular}
\end{table*}

\begin{table}[hbtp]
 \centering
  \caption{Hyperparameters of XGBoost on S1-S5}
  \label{tab:Parameters of XGBoost}
\begin{tabular}{cccccc}
\hline
              & S1     & S2     & S3     & S4     & S5     \\ \hline
max\_depth    & 1      & 1      & 1      & 3      & 3      \\
n\_estimaters & 3      & 3      & 3      & 3      & 3      \\
Acc           & 0.7109 & 0.7132 & 0.7153 & 0.7683 & 0.7662 \\ \hline
\end{tabular}
\end{table}

We investigate how different models affect the storage of graph structures. We chose XGBoost, Decision Tree, Naive Bayes, 1D-CNN, LSTM, and GRU models for comparison. In XGBoost, we set max\_depth at 5. Decision Tree and Naive Bayes are implemented in the standard sklearn library \cite{sklearn}. 1D-CNN refers to one-dimension Convolutional Neural Network (CNN), which is a variant of CNN. 1D-CNN, LSTM, and GRU are implemented in the torch.nn library \cite{torchnn}, with the batch size set at 4096, the epcoh at 3, and the learning rate at 0.001.

We look into the disk usage and the storage time cost of the graph structure as results shown in Figure \ref{differentmodels}. The XGBoost model spends the least amount of time storing the provenance graph structures while occupying the smallest disk storage. In addition, the machine learning model, XGboost, requires CPU for training, while 1D-CNN, LSTM, and GRU usually require GPU for training. LESS framework for different models compatibility is very good, the gap between different models is not very obvious. Switching to a different model when using LESS to store provenance graphs does not require any changes to the rest of the LESS framework, so it can be adjusted according to the actual hardware conditions.

\subsubsection{Provenance Graph Size}\label{graphsize}


In order to explore how LESS performs when different graph sizes change, we evaluate LESS on S1 to S5 datasets, where the sizes of the provenance graph and the raw log exhibit nearly linear growth. The window size is set at 6 for calculating the similarity matrix in the graph attribute storage. We measure the disk usage, storage time cost, and query speed as shown in Table \ref{tab:LESS Performance S1-S5}.


 The disk usage of the five datasets compressed by LESS increases almost linearly as the size of the input provenance graph increases linearly. 
The size of the model varies depending on the hyperparameters we choose. By tuning the two parameters max\_depth and n\_estimaters, we achieve almost consistent accuracy on the five datasets. The values of the hyperparameters and accuracy are shown in Table \ref{tab:Parameters of XGBoost}. The data in the table can be used as a reference for hyperparameters setting when the storage is executed.
In the query, the query speed on different datasets is basically consistent. The warm\_up time increases linearly because the size of the output to be predicted by the model and the size of the calibration table both increase linearly.
The storage usage of LESS on S1 to S5 is 7.0\% on average of the original provenance graph, which is larger than that of LESS on D1 to D5. This is because there is a significant difference between the two datasets, wherein the attribute strings in S1 to S5 are much longer and require a larger size of attribute tree for storage.

\section{Discussion}

LESS is a highly efficient storage system designed for provenance graphs, and it implements complimentary storage performance while maintains a time efficiency query of provenance graphs. Different from the existing database system, inserting, deleting, updating, and querying provenance graphs are not always necessary in LESS. Provenance graphs, often referred to as "the post-mortem forensic evidence", do not need modification after storage. Less supports the storage of an entire provenance graph and allows for the deletion of the compressed files. An integrity-maintained provenance graph can be queried directly from the disk. Additionally, LESS supports write-once-read-multiple-times tasks and does not guarantee ACID (Atomicity, Consistency, Isolation, Durability) for transaction. Provenance graphs stored in LESS do not lose any information and can be queried as the original complete version. Therefore, LESS is able to execute a sequence of operations on the stored provenance graphs, including graph reduction, anomaly detection, and attack investigation\cite{sok}. Additionally, LESS can store a reduced or labeled provenance graph without altering the storage format. In conclusion, LESS is orthogonal to the above works.

LESS supports forward-tracing and backward queries on provenance graphs. Operations on the provenance graphs are usually decided by downstream work, with which LESS is compatible because it can provide complete provenance graphs efficiently. LESS can be leveraged as the core component of a provenance graph management system with archival functions, but the other parts of the system still need to be further implemented, such as classification of the provenance graphs, access check, and hardware management. 

LESS provides an algorithm for storing graph structures based on machine learning models and demonstrates some advantages of the Xgboost model for this work. Extending LESS to the storage of other data may pose new challenges, with the design of the storage method being related to the query method, the data structure, and the domain characteristics of the data. Deep learning models are often considered to have better capabilities compared to machine learning models, but this is clearly not always justified in the storage domain. How learning models can be utilized to store a wide variety of data is a new direction in AI that still needs to be explored further in the future.

\section{Related Work}

{\bfseries Intrusion Detection. } Intrusion detection can be based on provenance graphs to determine the attack trail through automated analysis. Depending on the detection methods, intrusion detection mainly falls into two classic approaches, heuristic or rule-based and anomaly-based detection. Heuristic-based detection scheme exploits existing or expected knowledge of attack behaviors to define event matching rules, and matches the rules against audit logs or provenance graphs. If the match level reaches a certain threshold, it is considered an anomaly attack \cite{poirot, sleuth, ATT&CK}. HOLMES\cite{holmes} constructs provenance graph patterns based on information flow system, which are then matched against alerts to identify potential attacks. Hassan et al. propose Rapsheet\cite{rapsheet} and Nodoze\cite{nodoze} to analyze behaviors on provenance graphs, and the alerts are subsequently compared against a set of historical data or knowledge base to investigate threats. Anomaly-based detection defines a model of benign behavior on historical events. Deviations detected from the model of typical behaviors are considered anomalous\cite{frappuccino, pidas, deeplog, deepcase, pagoda, p-gaussian}. Modeling for provenance graphs is a key challenge to these anomaly-based approaches. Manzoor et al.'s Streamspot\cite{streamspot} transforms provenance graphs into fixed-length binary vectors by converting the nodes into multiple fixed-length chunks. Han et al. propose Unicorn\cite{unicorn} to construct a histogram description based on local provenance graphs and then hash it into fixed-length vectors. LESS can provide services to these detection methods, which need to manipulate the provenance subgraphs or the information flow of provenance graphs.

{\bfseries Attack Investigation. } 
Attack investigation further analyzes the detected alerts and diagnoses the root cause and consequences of attacks\cite{backtracking, propatrol, atlas, shadewatcher, provtalk}. Provenance analysis correlates and clusters the triggered alerts by similarity-based and causality-based techniques \cite{sok}.  Similarity-based methods derive alert correlation based on similarities shared by given alerts and historical threats and determine the severity of alerts\cite{hercule}. Causal analysis\cite{rapsheet} assesses whether the present alerts are related to historical threats. Alarms can be clustered by the frequency of historical events and assigned by a threat score for alert prioritization\cite{swift}. In addition to collecting system logs, application logs help better interpret system activities as user behaviors\cite{omegalog}. If application logs are associated with GUI elements, user actions can be more clearly defined\cite{uiscope}. Based on the provenance graph, AI can be used to infer user behaviors\cite{watson}, or group the activities represented by the logs into finer-grained communities to detect adversary behaviors\cite{beep}. Attack investigation requires a large number of logs over a long span and of multiple types. Therefore, the key challenges we face are how to store the voluminous size of logs and provenance graphs and execute effective queries.

{\bfseries Provenance Graph Reduction. } 
Provenance graph reduction algorithms have been proposed to compress large scale log data to optimize investigation tasks. The popular methods are semantic pruning, information flow preservation, and causal approximation\cite{sok}. LogGC\cite{loggc} categorizes the entries in the logs, and  performs Garbage Collecting to removes the less-important logs. NodeMerge\cite{nodemerge} combines nodes and edges of library files and combines associated with the temporary files into a single template node, with a single edge representing the reads to these files. Xu et al. merge provenance graph elements corresponding to events which share the same trackability\cite{cprpcar}. Hossain et al. propose full dependence preservation and source dependence preservation to achieve compression of logs through encoding and frequent field substitutions\cite{fdsd}. LogApprox \cite{logapprox} formally defines the \verb|"|forensic validity\verb|"| when dealing with the traceability of different types of attacks, and proposes three criteria, namely lossless, causality-preserving and attack-preserving. Hassan et al. achieve replacement compression of logs with the same pattern by constructing regular expressions for the logs behavioral patterns of multiple nodes within a cluster\cite{winnower}, reducing the overhead of storing and transmitting logs from a single node to a central server. FAuST\cite{faust} develops a unified log compression framework that allows modular additions and removals of the above methods for customized compression of logs. Depcomm\cite{depcomm} partitions a large provenance graph into process-entrance-centered centered subgraphs (communities) to generate a summary graph, eliminating unimportant or redundant edges within each subgraph, and generating a summary graph for each subgraph using InfoPath to represent the information flow of the subgraph. Compared to the reduction algorithm, LESS preserves the information in the provenance graph. Provenance graphs can also be reduced before or after storage, and LESS is compatible with these methods.

{\bfseries Provenance Graph Storage. }
Less information loss in compression or lossless compression optimizes downstream tasks to conduct effective investigation and forensic analysis. Fei et al.'s SEAL\cite{seal} method proposes a query-friendly compression (QFC) on the provenance graph and stores the original provenance graph with nearly all information preserved, that is, lossless compression. The SEAL method splits a whole provenance graph into two parts,graph structure and edge attributes, and compresses them separately. In querying, the compressed information is decompressed and returned to the original data. SEAL does not explore the storage challenges of provenance graphs. Alternatively, Ma et al. propose a learning model-based provenance graph storage method, LEONARD\cite{leonard}. LEONARD innovatively introduces DNNs to learn and represent graph data. LEONARD firstly processes the provenance graph into vertices and edges suitable for model training, then uses LSTM to learn the data and leverage a calibration table to correct mispredictions. Compared to these existing methods, LESS has better performance in store provenance graph while keeping low disk storage and effective query execution.

\section{Conclusion}
A fine-grained provenance graph is generated by log files, which collect host or network activities within a continuous period, enabling more effective intrusion detection and attack investigation. However, the high storage overhead of large size provenance graphs is of concern. In this paper, we propose a novel storage method for the provenance graph, LESS, to tackle the storage difficulties. In LESS, the provenance graph is creatively partitioned into two distinct components, namely the graph structure and attributes, which are stored separately. Based on their respective characteristics, we have devised appropriate storage schemes. We leverage a trained XGBoost model to store graph structures and use a tree structure to record the attributes of nodes and edges with less information redundancy. While maintaining lossless information, we support nearly real-time query. The experimental results show that LESS is efficient in terms of disk usage, storage time, query speed and memory usage.

\bibliographystyle{IEEEtranS}
\bibliography{sample-base}

\end{document}